\documentclass[prx,twocolumn,superscriptaddress,citeautoscript,amsart,longbibliography]{revtex4-1}

\usepackage{graphicx}
\usepackage{multirow}
\usepackage{color}
\usepackage{bm}
\usepackage{times}
\usepackage{amsmath,bm,amsfonts}
\usepackage{dcolumn}
\usepackage{graphicx}
\usepackage{latexsym}
\usepackage{braket}
\usepackage{bbold}

\def\bb{\mathbb}

\begin{document}
\title{
Higher-order topological superconductivity of spin-polarized fermions
}
\author{Junyeong \surname{Ahn}}
\author{Bohm-Jung\surname{Yang}}
\email{bjyang@snu.ac.kr}
\affiliation{Center for Correlated Electron Systems, Institute for Basic Science (IBS), Seoul 08826, Korea}
\affiliation{Department of Physics and Astronomy, Seoul National University, Seoul 08826, Korea}
\affiliation{Center for Theoretical Physics (CTP), Seoul National University, Seoul 08826, Korea}

\begin{abstract}
We study the superconductivity of spin-polarized electrons in centrosymmetric ferromagnetic metals. Due to the spin-polarization and the Fermi statistics of electrons, the superconducting pairing function naturally has odd parity. 
According to the parity formula proposed by Fu, Berg, and Sato, odd-parity pairing leads to conventional first-order topological superconductivity when a normal metal has an odd number of Fermi surfaces.
Here, we derive generalized parity formulae for the topological invariants characterizing higher-order topology of centrosymmetric superconductors. Based on the formulae, we systematically classify all possible band structures of ferromagnetic metals that can induce inversion-protected higher-order topological superconductivity. Among them, doped ferromagnetic nodal semimetals are identified as the most promising normal state platform for higher-order topological superconductivity. In two dimensions, we show that odd-parity pairing of doped Dirac semimetals induces a second-order topological superconductor. In three dimensions, odd-parity pairing of doped nodal line semimetals generates a nodal line topological superconductor with monopole charges. On the other hand, odd-parity pairing of doped monopole nodal line semimetals induces a three-dimensional third-order topological superconductor. Our theory shows that the combination of superconductivity and ferromagnetic nodal semimetals opens up a new avenue for future topological quantum computations using Majorana zero modes.
\end{abstract}


\maketitle

{\it Introduction.---}
Recently, odd-parity superconductivity has received great attention due to its potential to realize topological superconductors (TSCs)~\cite{sato2017topological,alicea2012new,sarma2015majorana,kitaev2003fault,nayak2008non}. Fu and Berg~\cite{fu2010odd}, and also Sato~\cite{sato2009topological,sato2010topological} proposed a simple but powerful parity formula relating the parity configuration in the normal state and the topological property of the odd-parity superconducting state. The simplicity of the formula allows a fast diagnosis of the topological nature of a superconducting state by just counting the number of Fermi surfaces, which greatly facilitates the search for TSCs in centrosymmetric materials.

One limitation of the Fu-Berg-Sato formula is that it can be applied only to conventional first-order TSCs in which $d$-dimensional bulk topology supports gapless Majorana states on $(d-1)$-dimensional boundaries. However, recent studies on topological crystalline phases have uncovered higher-order TSCs whose $d$-dimensional bulk topology protects gapless Majorana fermions on the boundaries with dimensions lower than $(d-1)$~\cite{khalaf2018higher,you2018higher,geier2018second,trifunovic2019higher,
wang2018weak,pahomi2019braiding}. 
In general, $k$th-order TSCs in $d$ dimensions host $(d-k)$-dimensional boundary Majorana states. In the case of $d$th-order TSCs in $d$ dimensions, Majorana zero modes (MZMs) exist at the corners, which can be potentially useful for topological quantum computations~\cite{alicea2012new,sarma2015majorana,kitaev2003fault,nayak2008non}.

Up to now, several interesting ideas have been proposed to realize 2D second-order TSCs in various different settings, such as using the superconducting proximity effect on quantum Hall insulators~\cite{liu2018majorana}, quantum spin Hall insulators~\cite{wang2018weak,wang2018high,yan2018majorana}, second-order topological insulators~\cite{hsu2018majorana}, Rashba semiconductors~\cite{zhu2018second} and nanowires~\cite{laubscher2019fractional}; breaking time reversal symmetry of TSCs with helical Majorana edge states by applying external magnetic field~\cite{khalaf2018higher,zhu2018tunable,volpez2019second,hsu2019inversion,wu2019plane} or attaching antiferromagnets~\cite{zhang2019higher}; and some other ideas~\cite{wang2018weak,wu2019higher,franca2019phase,hsu2019inversion,kheirkhah2019majorana}.
In 3D, on the other hand, there are only few mechanisms proposed for realizing a third-order TSC such as applying magnetic field to a 3D second-order TSC with helical hinge modes~\cite{khalaf2018higher}. For more systematic investigations of higher-order TSCs, a simple criterion for diagnosing higher-order band topology, similar to the Fu-Berg-Sato parity formula for first-order TSCs, is highly desired.
Although some formulae for higher-order TSCs having gapless boundary states were proposed recently~\cite{ono2019symmetry}, the parity formula for $d$th-order TSCs hosting MZMs is still lacking.

In this paper, we establish generalized parity formulae for higher-order TSCs and apply them to ferromagnetic metals where odd-parity superconductivity naturally arises. Using the generalized parity formulae, we classify all possible spin-polarized band structures of  centrosymmetric ferromagnetic metals that can realize inversion-protected higher-order TSC. 
From this analysis, we find doped ferromagnetic nodal semimetals as an ideal normal state that realizes higher-order TSCs.
Explicitly, in 2D, odd-parity pairing of a doped Dirac semimetal (DSM) induces a 2D second-order TSC. In 3D, odd-parity pairing of a doped nodal line semimetal (NLSM) generates a nodal line superconductor with monopole charges. 
Furthermore, in the case of a doped monopole NLSM~\cite{fang2015new,ahn2018band}, odd-parity pairing induces a 3D third-order TSC.
These findings show that the combination of superconductivity and spin-polarized 2D and 3D nodal semimetals can be promising platforms for topological quantum computations using MZMs.

{\it Symmetry and nodal structures.---}
Let us first clarify the symmetry of the normal and superconducting states of ferromagnetic metals with inversion symmetry $P_0$ and classify the relevant nodal structures.
We assume that an electron's spin is polarized along the $z$-direction.
Also, we neglect spin-orbit coupling, but its influence is discussed later.
In this setting, although time reversal symmetry ${\cal T}=i\sigma_y K$ is broken, the ferromagnetic metallic state is symmetric under
the effective time reversal $T\equiv e^{i\pi \sigma_y/2}{\cal T}=K$ defined as the product of ${\cal T}$ and a $180^{\circ}$ spin rotation around the $y$ axis, $e^{i\pi \sigma_y/2}$.
Here $\sigma_y$ is a Pauli matrix for spin degrees of freedom, and $K$ denotes the complex conjugation operator. 
Also, $P_0=P_0^*$ because $[P_0,T]=0$.
Then, the system is invariant, locally at each momentum ${\bf k}$, under $P_0T$ symmetry satisfying $(P_0T)^2=1$. Such a $P_0T$ symmetric system belongs to the ${\bf k}$-local symmetry class AI$+{\cal I}$ proposed by Bzdusek and Sigrist~\cite{bzdusek2017robust}, where the 1D and 2D topological phases are classified by $Z_2$ invariants~\cite{fang2015new,bzdusek2017robust}.
Here the 1D $Z_2$ invariant is the quantized Berry phase, which is the topological charge of 2D Dirac points and also of 3D nodal lines. The 2D $Z_2$ invariant is the monopole charge of 3D nodal lines.

To describe the superconducting state, we introduce a $2N$-component Nambu spinor $\hat{\Psi}({\bf k})=[\hat{c}_{\uparrow\alpha}({\bf k}),\hat{c}^{\dagger}_{\uparrow\beta}({\bf k})]^T$, where $\hat{c}_{\uparrow\alpha}({\bf k})$ [$\hat{c}^{\dagger}_{\uparrow\alpha}({\bf k})$] is an electron creation [annihilation] operator with spin up and the orbital indx $\alpha=1,\hdots,N$. 
The corresponding Bogoliubov-de Gennes (BdG) Hamiltonian can be written as
$
\hat{H}
=\hat{\Psi}^{\dagger}H_{\rm BdG}\hat{\Psi},
$
where
\begin{align}\label{eq:HBdG}
H_{\rm BdG}
=
\begin{pmatrix}
h({\bf k})
& \Delta({\bf k})\\
\Delta^{\dagger}({\bf k})
&-h^T(-{\bf k})
\end{pmatrix}.
\end{align}
Here, $h({\bf k})$ indicates the Hamiltonian for the normal state, and
the pairing function $\Delta_{\alpha\beta}({\bf k})$ with orbital indices $\alpha,~\beta$ satisfies
$\Delta_{\alpha\beta}({\bf k})=-\Delta_{\beta\alpha}(-{\bf k})$ because of the Fermi statistics of electrons.
Since the pairing function forms an irreducible representation of the symmetry group, it can have either odd-parity $P_0\Delta({\bf k})P_0^{-1}=-\Delta(-{\bf k})$ or even-parity $P_0\Delta({\bf k})P_0^{-1}=+\Delta(-{\bf k})$.

In the weak-pairing limit, we can focus on the pairing at the Fermi energy $E_F$ and define the corresponding pairing function as $\Delta_{E_F}({\bf k})$.
Then, $P_0\Delta_{E_F}({\bf k})P_0^{-1}=\Delta_{E_F}({\bf k})$ because $\Delta_{E_F}$ is a $1\times 1$ matrix.
The Fermi statistics $\Delta_{E_F}({\bf k})=-\Delta_{E_F}(-{\bf k})$ naturally shows that 
the pairing function satisfies the odd-parity condition
\begin{align}
\label{eq:odd-parity}
P_0\Delta_{E_F}({\bf k})P_0^{-1}=-\Delta_{E_F}(-{\bf k}).
\end{align}
Therefore, in Eq.~(\ref{eq:HBdG}), we consider only odd-parity pairing functions that satisfy $P_0\Delta({\bf k})P_0^{-1}=-\Delta(-{\bf k})$ (See also the Supplemental Material~\cite{supp}).
The corresponding odd-parity BdG Hamiltonian is symmetric under inversion $P=\tau_z P_0$ which anticommutes with the particle-hole symmetry $C=\tau_x K$, where $\tau_{x,y,z}$ are Pauli matrices for the Nambu space.
$PT$ and $CP$ symmetries satisfying $(PT)^2=1$ and $(CP)^2=-1$, which show that the BdG Hamiltonian belongs to the ${\bf k}$-local symmetry class CI$+{\cal I}$~\cite{bzdusek2017robust}.
In this class, 2D Dirac points or 3D nodal lines can be protected as in the case of the class AI$+{\cal I}$.
The only difference is that the 1D invariant is integer-valued in the class CI$+{\cal I}$, but this is irrelevant in our analysis below because we are only interested in the parity of the 1D invariant that can be related to the eigenvalues of $P$.

\begin{figure}[t!]
\includegraphics[width=8.5cm]{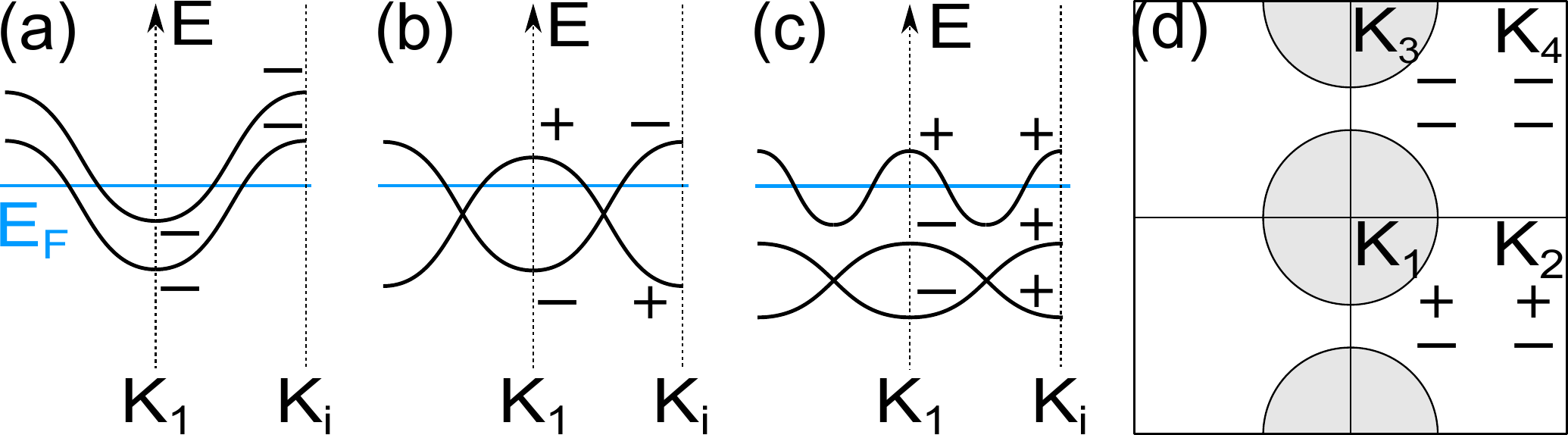}
\caption{Band structure and parity configuration of spin-polarized metals leading to 2D second-order TSCs in the weak pairing limit.
(a) Two electron-like (or hole-like) Fermi surfaces surrounding the same TRIM. 
(b) Doped DSM with $\nu_1=1$.
(c) Normal state whose whole bands, including both occupied and unoccupied bands, have the higher-order topology with $\nu_2=1$.
The horizontal axes in (a,b,c) schematically represent the 2D Brillouin zone:
${\bf K}_1=(0,0)$, and ${\bf K}_i$ indicates the other three TRIMs with the same parity configuration.
$\pm$ represents the parity at TRIM.
(d) The fourth way to obtain the higher-order TSCs.
Here, the $\pm$ sign on the top (bottom) row at each TRIM represents the parity of the higher-energy (lower-energy) states.
One (no) band is occupied in the gray (white) regions, and the boundaries show the relevant Fermi surfaces.
}
\label{fig:mechanism}
\end{figure}


{\it Nodal structure of TSC and parity formula.---}
According to Eq.~\eqref{eq:odd-parity}, an odd-parity pairing function $\Delta({\bf k})$ changes its sign on the Fermi surfaces 
surrounding a time-reversal-invariant momentum (TRIM) so that an even number of nodes should appear at the points where the sign of $\Delta({\bf k})$ changes.
The number of nodal points can be related with the inversion parities of occupied bands using the idea proposed in Refs.~\cite{fu2010odd,sato2009topological,sato2010topological} as follows.
In 2D, the parity of the number of Dirac node pairs related by inversion can be counted by the $Z_2$ invariant $\nu_1\equiv\sum_{{\bf K}\in {\rm TRIM}}
n^{\rm o}_-({\bf K}) \mod 2$~\cite{kim2015dirac}, where $n^{\rm o}_-({\bf K})$ is the number of occupied states with negative parity at ${\bf K}$. Here $\nu_1$ can be understood as the number of band inversions at TRIM that create pairs of Dirac points, starting from the trivial phase with only positive-parity occupied states.

One can define a similar parity index $\nu_1^{\rm BdG}$ for the BdG Hamiltonian as
\begin{align}
\label{eq:nu1}
\nu_1^{\rm BdG}
&\equiv
\sum_{{\bf K}\in {\rm TRIM}}
n^{\rm BdG;o}_-({\bf K})
\notag\\
&=\sum_{{\bf K}\in {\rm TRIM}}
n^{\rm o}_-({\bf K})+n^{\rm u}_+({\bf K})\notag\\
&=\sum_{{\bf K}\in {\rm TRIM}}
n^{\rm u}({\bf K}) \mod 2,
\end{align}
where $n^{\rm o (u)}_{\pm}$ is the number of occupied (unoccupied) states with $\pm$ parity in the normal state, $n^{\rm u}=n^{\rm u}_++n^{\rm u}_-$, and $n^{\rm BdG; o(u)}_{\pm}$ is defined similarly for the BdG Hamiltonian with an odd-parity pairing function.
The second line in Eq.~(\ref{eq:nu1}) results from the odd-parity pairing, and the third line follows from $n^{\rm o}_-({\bf K})=n^{\rm o}_-({\bf K})+n^{\rm u}_-({\bf K})-n^{\rm u}_-({\bf K})=n^{\rm o}_-({\bf K})+n^{\rm u}_-({\bf K})+n^{\rm u}_-({\bf K}) \mod 2$ together with $\sum_{\bf K}n_-({\bf K})=0 \mod 2$ following from that,  when all the bands are occupied, no band crossing exists at the Fermi level.
Equation~\eqref{eq:nu1} shows that $\nu^{\rm BdG}_1=1 \mod 2$ only when there exists an odd number of Fermi surfaces.
This is consistent with the odd-parity condition of the pairing function $\Delta({\bf k})$ on the Fermi surface in Eq.~(\ref{eq:odd-parity}), which guarantees an odd number of Dirac node pairs in the superconducting state per each normal state Fermi surface enclosing a TRIM.

{\it Generalized parity formula for second-order TSC in 2D.---}
To derive the condition for higher-order superconductivity of spin-polarized electrons, let us introduce generalized parity formulae.
According to the Dirac Hamiltonian formalism for inversion-protected higher-order topological phases~\cite{khalaf2018higher,hwang2019fragile}, we can obtain a higher-order TI by inverting $2^n$ bands at a TRIM starting from a topologically trivial phase. Here, $n$ denotes a nonnegative integer.
Therefore, counting the number of the simultaneous inversion of $2^n$ bands at TRIM leads to the following $Z_2$ index,
\begin{align}
\label{eq:nun}
\nu_{2^n}\equiv\sum_{{\bf K}\in {\rm TRIM}}\left[\frac{n^{\rm o}_-({\bf K})}{2^n}\right]_{\rm floor} \mod 2,
\end{align}
where $\left[m+a\right]_{\rm floor}=m$ for an integer $m$ and $0\le a<1$.
We can also introduce similar indices $\nu^{\rm BdG}_{2^n}$ for the BdG Hamiltonian by replacing $n^{\rm o}_-({\bf K})$ by $n^{\rm BdG;o}_-({\bf K})$.
These indices characterize higher-order TSCs.

Let us first discuss the physical meaning of $\nu^{\rm BdG}_2$ in 2D. 
Recently, it was shown that $\nu_2=1$ indicates the second-order topology of a $PT$-symmetric topological insulator with chiral symmetry, characterized by fractional corner charges on the boundary~\cite{wang2018higher,ahn2019failure,hwang2019fragile}. A straightforward extension of this idea shows that $\nu^{\rm BdG}_2=1$ characterizes a second-order TSC with Majorana corner modes.
Explicitly, $\nu^{\rm BdG}_2$ can be decomposed as
\begin{align}
\label{eq:nu2}
\nu^{\rm BdG}_2
&=
\sum_{{\bf K}\in {\rm TRIM}}
\left[
\frac{n^{\rm u}({\bf K})}{2}
\right]_{\rm floor}
+
\sum_{{\bf K}\in {\rm TRIM}}
n^{\rm o}_-({\bf K})
\notag\\
&
+
\sum_{{\bf K}\in {\rm TRIM}}
\left[
\frac{n_-({\bf K})}{2}
\right]_{\rm floor}
+
\sum_{{\bf K}\in {\rm TRIM}}
\delta_2({\bf K}) \mod 2,
\end{align}
where $\delta_2({\bf K})=[n^{\rm u}({\bf K})+1]n_-({\bf K}) \mod 2$. The detailed derivation is in the Supplemental Material~\cite{supp}. 
In Eq.~(\ref{eq:nu2}), the first term counts the parity of the number of ``double Fermi surfaces", that is, two electron-like (or hole-like) Fermi surfaces enclosing the same TRIM, in the normal state. The second term is $\nu_1$ for the occupied bands in the normal state and the third term is $\nu_2$ when all bands are occupied in the normal state. Finally, the last term counts the number of TRIM with an even number of unoccupied state and an odd number of negative-parity eigenstates.
Figures~\ref{fig:mechanism}(a-d) show four different normal state band structures leading to $\nu^{\rm BdG}_2=1$ in the weak-pairing limit, which arise from the nontrivial value of the first, second, third, and fourth terms in Eq.~\eqref{eq:nu2}, respectively.

The analysis of Eq.~(\ref{eq:nu2}) becomes much simpler in systems with an inversion-symmetric unit cell, where all atoms in a unit cell can be adiabatically shifted to its center without breaking inversion symmetry. In this case, the third term in Eq.~\eqref{eq:nu2} vanishes because an inversion-symmetric unit cell gives a topologically trivial state with $\nu_2=1$ when all bands are occupied.
Similary, the zero Berry phase of the whole bands makes the fourth term vanish (See Supplemental Material~\cite{supp}).


Then, there remain two different channels leading to $\nu^{\rm BdG}_2=1$: One is odd-parity pairing in a metal with double Fermi surfaces, and the other is odd-parity pairing in a doped DSM, whose nontrivial band topology arises from the first and second terms in Eq.~\eqref{eq:nu2}, respectively.
In general, the former induces nodal superconductivity rather than a fully gapped TSC.
This is because each of the two Fermi surfaces encloses a TRIM so that an odd-parity pairing function accompanies the sign reversal at two points on the Fermi surface, generating Dirac nodes.
A strong pairing is required to get a fully gapped superconducting state via pair annihilations of Dirac nodes, unless the system is fine-tuned so that the two Fermi surfaces are very close to each other.
On the other hand, even weak pairing generates a fully gapped superconducting state in doped DSMs because two disconnected Fermi surfaces, each centered at a generic momentum, are paired in this case.

{\it Higher-order TSCs in 3D and further generalization.---}
In 3D, $\nu_1=1$ indicates an odd number of nodal lines~\cite{kim2015dirac}, and
$\nu_2=1$ indicates an odd number of pairs of monopole nodal lines in the Brillouin zone~\cite{ahn2018band,song2018diagnosis}.
Similarly, $\nu^{\rm BdG}_1=1$ ($\nu^{\rm BdG}_2=1$) indicates a superconductor with an odd number of nodal lines (monopole nodal line pairs).
In particular, the superconductor with a monopole nodal line pair exhibits the second-order topological property and carries anomalous hinge Majorana states, as in the case of chiral-symmetric monopole NLSMs~\cite{wang2018higher}.
Similar to 2D cases, the most promising way to get $\nu^{\rm BdG}_2=1$ is the process with a nontrival second term in Eq.~(\ref{eq:nu2}), 
which corresponds to doping spin-polarized NLSMs.
The third term in Eq.~\eqref{eq:nu2} always vanishes when the whole bands are fully considered. 
Also the fourth term vanishes if we take an inversion-symmetric unit cell as in 2D.
In the case of the first term, it may be relevant in a strong pairing limit. A double Fermi surface normally generates a superconducting state with nodal lines carrying trivial monopole charges from each Fermi surface.
When the pairing amplitude is sufficiently strong, however, the two trivial nodal lines may recombine and turn into two monopole nodal lines.
We note that the same mechanism corresponding to the second term in Eq.~\eqref{eq:nu2} was also proposed in Ref.~\cite{bzdusek2017robust} for systems with SU(2) spin rotation symmetry.

The above formulation can be generalized further to $\nu^{\rm BdG}_{2^n}$ with an arbitrary $n$:
\begin{align}
\label{eq:nu2n}
\nu^{\rm BdG}_{2^n}
&=
\sum_{{\bf K}\in {\rm TRIM}}
\left[
\frac{n^{\rm u}({\bf K})}{2^n}
\right]_{\rm floor}
+
\sum_{{\bf K}\in {\rm TRIM}}
\left[
\frac{n^{\rm o}_-({\bf K})}{2^{n-1}}
\right]_{\rm floor}
\notag\\
&
+
\sum_{{\bf K}\in {\rm TRIM}}
\left[
\frac{n_-({\bf K})}{2^n}
\right]_{\rm floor}
+
\sum_{{\bf K}\in {\rm TRIM}}
\delta_{2^n}({\bf K})\mod 2,
\end{align}
where the definition of $\delta_{2^n}({\bf K})$ is given in the Supplemental Material~\cite{supp}.
In particular, $\nu^{\rm BdG}_4=1$ characterizes the third-order TSC in 3D~\cite{hwang2019fragile}.
By the same reason discussed above, one can show that the best way to get a fully gapped superconductivity with $\nu^{\rm BdG}_{4}=1$ is to use the process related with the second term in Eq.~(\ref{eq:nu2n}), which can be achieved by doping a monopole NLSM (see the Supplemental Material for details~\cite{supp}). To sum up, in ferromagnetic systems with an inversion-symmetric unit cell, doped nodal semimetals are the best normal state to get a higher-order TSC in the weak-pairing limit.

\begin{figure}[t!]
\includegraphics[width=8.5cm]{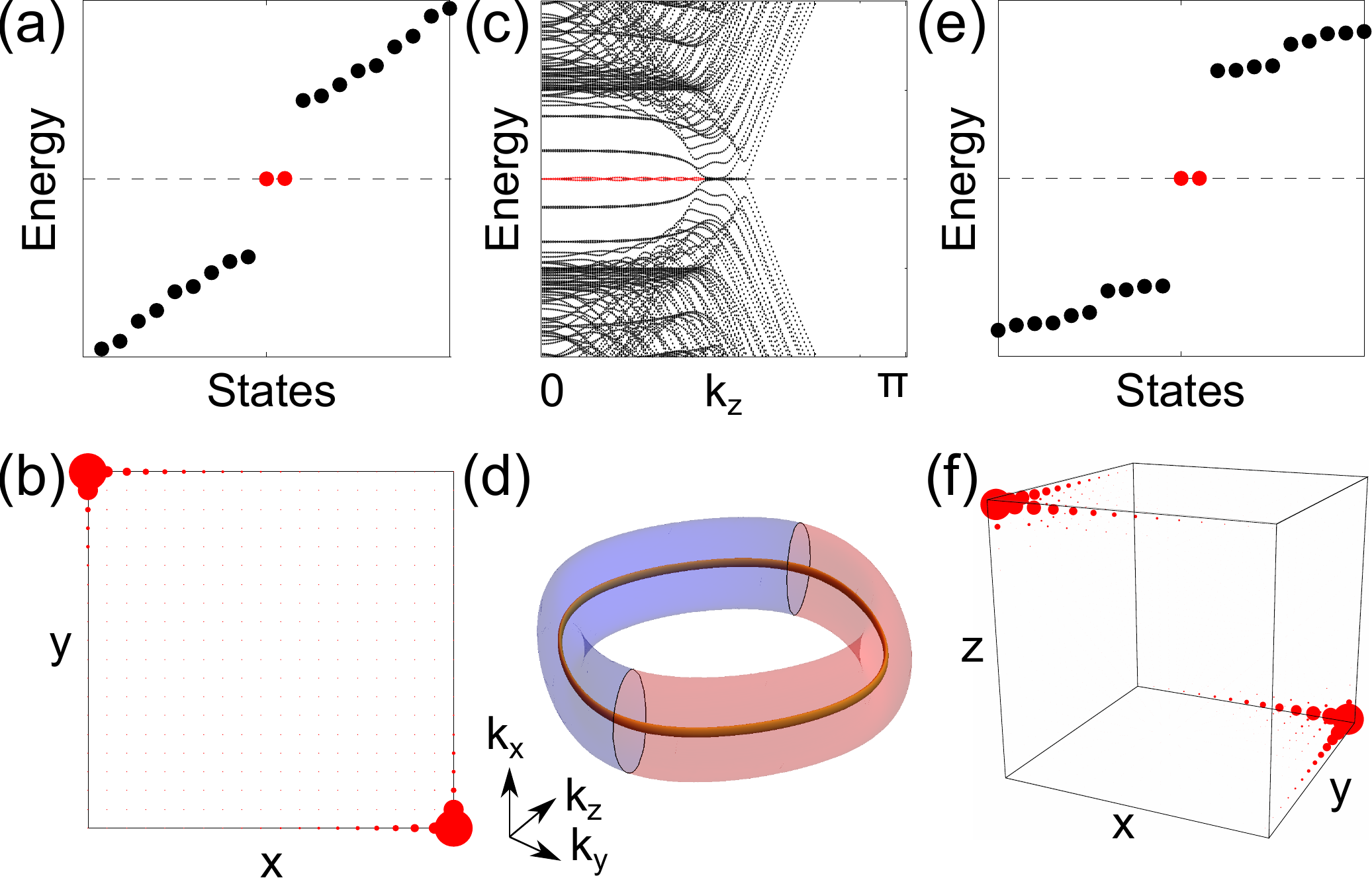}
\caption{
Higher-order topological superconductivity from lattice models.
(a,b) 2D second-order TSC obtained by adding an odd-parity pairing function to the doped 2D DSM described in Eq.~\eqref{eq:2dSM}.
(a) Energy spectrum on a finite-size lattice.
(b) Probability density of a Majorana zero mode 
(c,d) Monopole nodal line superconductor derived from a doped 3D NLSM. 
(c) Energy spectrum of the system, finite-sized along $x$ and $y$ directions.
(d) Nodal structure in the Brillouin zone.
The torus indicates the Fermi surface enclosing a nodal line (thick gold line) in the normal state. The blue (red) color indicates the region where the pairing function has positive (negative) sign. Two monopole nodal loops appear at the interection, where the pairing function changes its sign.
(e,f) 3D third-order TSC  derived from a doped 3D monopole NLSM
(e) Energy spectrum on a finite-size lattice.
(f) Probability density of a Majorana zero mode
}
\label{fig:higher-order}
\end{figure}

{\it Lattice model.---}
We demonstrate our theory by using simple tight-binding models defined on rectangular or orthorombic lattices.
We construct three models in which the spin-polarized normal states are a 2D DSM, a 3D NLSM, and a 3D monopole NLSM, respectively. When an odd-parity superconducting pairing is introduced, we show that the three nodal semimetals turn into a 2D second-order TSC, a 3D monopole nodal line superconductor, and a 3D third-order TSC, respectively.

First, a 2D DSM can be described by the nearest-neighbor tight-binding Hamiltonian for $s$ and $p_x$ orbitals as
\begin{align}
\label{eq:2dSM}
h
&=-\mu+2t\sin k_x\sigma_y+(M-2t\cos k_x-2t\cos k_y)\sigma_z,
\end{align}
where the Pauli matrices $\sigma_{y,z}$ describe the orbital degrees of freedom with $\uparrow$ ($\downarrow$) indicating a $s$ ($p_x$) orbital. 
The corresponding band structure exhibits two Dirac points on the $k_x=0$ line when $0<M/t<4$ at the energy $E=-\mu$.
To induce superconductivity, we consider the following interaction term 
$H_{\rm int}=-U\sum_{i,\sigma\ne\sigma'}n_{i,\sigma}n_{i,\sigma'}-V\sum_{\braket{i,j},\sigma}n_{i,\sigma}n_{j,\sigma}$
where $U$ ($V$) indicates the on-site interorbital (nearest-neighbor intraorbital) interaction, which is to be treated by mean-field approximation. 
The resulting odd-parity pairing leads to a fully gapped TSC whose second order band topology is clearly demonstrated in Fig.~\ref{fig:higher-order}(a,b). Vetically stacking the 2D DSM and introducing interlayer hopping, described by $-2t\cos k_z\sigma_z$, we obtain the Hamiltonian for a 3D NLSM. 
Also, by further adding $p_y$ and $d_{xy}$ orbitals at each lattice site and introducing nearest-neighbor hopping, we obtain a 3D monopole NLSM. 
Adding an odd-parity pairing function in these NLSMs leads to a 3D monopole nodal line superconductor and a 3D third-order TSC whose topological properties are demonstrated in Fig.~\ref{fig:higher-order}(c-f). Detailed information about the tight-binding models is given in the Supplemental Material~\cite{supp}.

{\it Discussions.---}
We first discuss the effect of the inversion asymmetry of the unit cell.
For instance, in the Kagome lattice, the unit cell always breaks inversion symmetry if all atoms are required to be strictly within the unit cell.
One may choose a unit cell, invariant under inversion up to lattice translations, only when the atoms in a unit cell are located on its boundary.
In this case, $\nu_2=1$ when each atom is occupied by one electron, so the third term in Eq.~\eqref{eq:nu2} is nontrivial ~\cite{supp} for a three-band tight-binding model.
Then, we have $\nu_2^{\rm BdG}=0$ even when the normal state is a doped DSM.
However, this does not mean that MZM is absent on the boundary.
In fact, one can show that MZMs exist (do not exist) when $\nu_2^{\rm BdG}=0$ ($\nu_2^{\rm BdG}=1$) in constrast to systems having an inversion-symmetric unit cell.
To obtain more conventional bulk-boundary correspondence where $\nu_2^{\rm BdG}=1$ 
always indicates the existence of MZMs independent of symmetry of the unit cell, 
one may define a reference trivial phase of the TSC as the limit $\mu\rightarrow -\infty$ where all electrons are unoccupied in the normal state as proposed in~\cite{skurativska2019atomic}.
This gives a well-defined trivial phase for TSCs because Majorana fermions are confined to form electrons in such a limit: $|\mu|$ serves as the binding energy for Majorana fermions because $\mu c^{\dagger}_{\bf x}c_{\bf x}=2i\mu\gamma_{1\bf x}\gamma_{2\bf x}$ at each site ${\bf x}$~\cite{kitaev2001unpaired}, where Majorana operators $\gamma_{1,2}$ are defined from the electron annihilation operator $c_{\bf x}=\gamma_{1\bf x}+i\gamma_{2\bf x}$.

Next, let us discuss the effect of spin-orbit coupling.
When spin-orbit coupling is included, $T$ symmetry is broken because the electron's spin cannot rotate freely independent of the orbital degrees of freedom.
Since the protection of the nodal structures in both normal and superconducting states requires the combination of time reversal and inversion symmetries, the nodal structures become unstable when spin-orbit coupling exists.
However, our formula in Eq.~\eqref{eq:nu2n} is still applicable as long as inversion symmetry is preserved.
Accordingly, a gapped higher-order TSC can still survive if the parity configuration does not change due to spin orbit coupling, since their topology can be protected by inversion symmetry only.
In the case of the monopole nodal line superconductor, the nodes are fully gapped when $T$ symmetry is broken due to spin-orbit coupling.
The resulting gapped superconductor is a second-order TSC hosting chiral hinge states~\cite{khalaf2018higher,khalaf2018symmetry,ahn2018band}.
In fact, in the normal state, NLSM transforms to a Weyl semimetal by spin-orbit coupling as long as the parity configuration does not change.
This means that, when spin-orbit coupling exists, what we observe is the transition from a Weyl semimetal to a fully gapped second-order TSC.

There exist many candidate materials for ferromagnetic spin-polarized 2D Dirac semimetals and some 3D nodal line semimetals~\cite{wang2018recent,zou2019study,wang2018nodal,kim2018large,chen2019weyl,zuo2019fully}.
One way to realize spin-triplet pairing in 2D ferromagnetic nodal semimetals is to use a superconductor-ferromagnet-superconductor heterostructure with inversion symmetry.
Here, we can use conventional spin-singlet $s$-wave superconductors and a ferromagnet with in-plane magnetization.
After spin-singlet Cooper pairs penetrate into the ferromagnet, they can turn into spin-triplet Cooper pairs because of the spin polarization in the ferromagnet~\cite{eschrig2003theory}.
In 3D, on the other hand, an intrinsic superconducting pairing is required because the proximity effect is not effective.
In fact, there are several materials where the coexistence of ferromagnetism and superconductivity is reported
including, uranium-based materials UGe$_2$, URhGe, UCoGe, UTe$_2$~\cite{aoki2019review,saxena2000superconductivity,huxley2001uge,aoki2001coexistence,huy2007superconductivity,huxley2015ferromagnetic,ran2019nearly}, and the more recently proposed
twisted double bilayer graphene~\cite{liu2019spin,lee2019theory,shen2019observation}.
We hope that our work stimulates the research on higher-order TSC in ferromagnets.
This will open a new route to Majorana quantum computations, where
ferromagnetic nodal semimetals with spin-polarized band crossing serve as platforms for higher-order TSCs.

Finally, let us briefly comment on the extension of our result to other symmetry classes.
We note that our parity formula Eq.~\eqref{eq:nun} is generally applicable to any odd-parity superconductors, while we focus on ferromagnetic systems with effective time reversal symmetry since odd-parity pairing is natural in these systems.
Furthermore, we expect that nodal semimetals required by eigenvalues of symmetry operator $G$ can lead to a $G$-protected $d$th-order TSCs, which can be shown by extending Eq.~\eqref{eq:nun} to eigenvalues of $G$, as is outlined in~\cite{ono2019symmetry} for $k$th-order TSCs with $k<d$. 
We leave more detailed theoretical analysis for future work.

\begin{acknowledgments}
We are grateful to A. Skurativska, T. Neupert, and M. H. Fischer for sharing a related manuscript~\cite{skurativska2019atomic}.
We thank Seunghun Lee, Yoonseok Hwang, Taekoo Oh, and Se Young Park for helpful discussions.
J.A. was supported by IBS-R009-D1.
B.-J.Y. was supported by the Institute for Basic Science in Korea (Grant No. IBS-R009-D1) and Basic Science Research Program through the National Research Foundation of Korea (NRF) (Grant No. 0426-20190008), and  the POSCO Science Fellowship of POSCO TJ Park Foundation (No. 0426-20180002).
This work was supported in part by the U.S. Army Research Office under Grant Number W911NF-18-1-0137.

{\it Note added.---}
Recently, we became aware of two related manuscripts~\cite{yan2019higher,skurativska2019atomic}. Reference~\cite{yan2019higher} also suggests that a third-order topological superconductivity can be obtained by odd-parity pairing in a doped nodal line semimetal.
In Ref.~\cite{skurativska2019atomic}, the authors also have related the higher-order topology of inversion-symmetric superconductors to the parity of the normal state.
\end{acknowledgments}

\clearpage
\newpage

\onecolumngrid

\begin{center}
\textbf{\large Supplemental Material for ``Higher-order topological ferromagnetic superconductivity"}
\end{center}

\begin{center}
Junyeong Ahn$^{1,2,3}$ and Bohm-Jung Yang$^{1,2,3,*}$\\
\vspace{0.1in}
{\small \it
$^1$Center for Correlated Electron Systems, Institute for Basic Science (IBS), Seoul 08826, Korea\\
$^2$Department of Physics and Astronomy, Seoul National University, Seoul 08826, Korea\\
$^3$Center for Theoretical Physics (CTP), Seoul National University, Seoul 08826, Korea}
\end{center}

\setcounter{section}{0}
\setcounter{equation}{0}
\setcounter{figure}{0}
\setcounter{table}{0}
\setcounter{page}{1}

\tableofcontents

\section{Odd-Parity Pairing in the Weak Pairing Limit in View of Symmetry and Topology}

\begin{table}[b]
		\caption{Classification of topological charges of nodes in spin-polarized centrosymmetric normal and superconducting states.
We adopt the notation proposed by Bzdusek-Sigrist~\cite{bzdusek2017robust} to indicate the nodal class.
Here, $P$, $T$, $C$, and $S$ are spatial inversion, time reversal, particle-hole, and chiral symmetry operators.
Since $P$, $T$, and $C$ all reverse the sign of the crystal momentum, $PT$, $PC$, and $S$ do not change the crystal momentum.
These are relevant for the protection of band degeneracies appearing at generic low-symmetry momenta in the Brillouin zone.
``0" in the entry means the absence of symmetry in the third-fifth column and means the absence of topologically nontrivial phase in the sixth-ninth column.
$\delta$ is the dimension of the submanifold enclosing a node in the Brillouin zone, on which the topological charge is defined.
The name in the parenthesis after ${\bb Z}$ or ${\bb Z}_2$ indicates the relevant topological invariant.}
		\begin{tabular}{c|c|ccc|ccc}
			\hline
			\hline
			Pairing-parity & Nodal class & $(PT)^2$ & $(PC)^2$ & $S^2$ & $\delta=0$ & $\delta=1$ & $\delta=2$ 
\\ \hline\hline
			Normal state & A$+{\cal I}$ & $0$ & $0$ & $0$ & $0$ & $0$ & $\mathbb{Z}$ (Chern number) 
\\
			Even & D$+{\cal I}$ & $0$ & $1$ & $0$ & $\mathbb{Z}_2$ (Pfaffian) & $0$ & $2\mathbb{Z}$ (Chern number) 
\\
			Odd & C$+{\cal I}$ & $0$ & $-1$ & $0$ & $0$  & $0$ & $\mathbb{Z}$ (Chern number) 
\\
\hline\hline
			Normal state & AI$+{\cal I}$ & $1$ & $0$ & $0$ & $0$ & $\mathbb{Z}_2$ (Berry phase) & $\mathbb{Z}_2$ (Stiefel-Whitney number) 
\\
			Even & BDI$+{\cal I}$ & $1$ & $1$ & $1$ & $\mathbb{Z}_2$ (Pfaffian) & $\mathbb{Z}_2$ (winding number) & $0$ 
\\
			Odd & CI$+{\cal I}$ & $1$ & $-1$ & $1$ & $0$ & $\mathbb{Z}$ (winding number) & $\mathbb{Z}_2$ (Stiefel-Whitney number) 
\\
			\hline
		\end{tabular}
		\label{tab:nodal}
\end{table}

Spin-polarized electrons in centrosymmetric systems form odd-parity pairing in the weak pairing limit, which can be understood from the Fermi statistics of electrons.
This can be alternatively explained from the viewpoint of topological charges of band crossing nodes.
Let us note that a nondegenerate Fermi surface in the normal state is described as a two-fold degenerate $(d-1)$-dimensional node in $d$ dimensions in the Bogoliubov-de Gennes (BdG) formulation in the absence of superconducting pairing.
Table~\ref{tab:nodal} shows that a zero-dimensional $Z_2$ topological charge exists in the superconducting phase (does not exist) for an even-parity (odd-parity) pairing, regardless of whether there exists effective time reversal symmetry under $T$ satisfying $T^2=1$.
If we consider even-parity pairing, the zero-dimensional $Z_2$ topological charge serves as the topological charge of the $(d-1)$-dimensional node.
Therefore, any weak even-parity superconducting pairing preserves the $(d-1)$-dimensional node in the BdG Hamiltonian.
To open the gap, two Fermi surfaces should meet to cancel the topological charge, and this process requires a strong superconducting pairing.
In contrast, weak odd-parity pairing opens the gap on the Fermi surface because $(d-1)$-dimensional node is not protected in the BdG formulation.
Accordingly, an odd-parity pairing state is energetically favorable than an even-parity pairing state.

\section{Parity Eigenvalues and Odd-Parity Pairing}

\subsection{Parity indices of odd-parity superconductors.}
In the main text, we define parity indices $\nu_{2^n}$ that counts the number of $2^n$ band inversion occuring at each time-reversal-invariant momentum (TRIM):
\begin{align}
\nu_{2^n}
=
\sum_{{\bf K}\in {\rm TRIM}}
\left[
\frac{n^{\rm o}_-({\bf K})}{2^n}
\right]_{\rm floor},
\end{align}
where we define $n^{\rm o (u)}_{\pm}({\bf K})$ as the number of occupied (unoccupied) states at ${\bf K}$ with inversion parity $\pm 1$.
One of our main result is the decomposition of the parity indices for the odd-parity BdG Hamiltonian into
\begin{align}
\label{eq:decomposition}
\nu^{\rm BdG}_{2^n}
=
\sum_{{\bf K}\in {\rm TRIM}}
\left[
\frac{n^{\rm u}({\bf K})}{2^n}
\right]_{\rm floor}
+
\sum_{{\bf K}\in {\rm TRIM}}
\left[
\frac{n^{\rm o}_-({\bf K})}{2^{n-1}}
\right]_{\rm floor}
+
\sum_{{\bf K}\in {\rm TRIM}}
\left[
\frac{n_-({\bf K})}{2^n}
\right]_{\rm floor}
+
\sum_{{\bf K}\in {\rm TRIM}}
\delta_{2^n}({\bf K})
\end{align}
modulo two, where $n^{\rm u}({\bf K})=n^{\rm u}_+({\bf K})+n^{\rm u}_-({\bf K})$, $n_-({\bf K})=n^{\rm o}_-({\bf K})+n^{\rm u}_-({\bf K})$, and $\delta_{2^n}({\bf K})$ is defined below.
Here, we derive the decomposition Eq.~\eqref{eq:decomposition} as follows:
\begin{align}
\nu^{\rm BdG}_{2^n}
&=\sum_{{\bf K}\in {\rm TRIM}}\left[\frac{n^{\rm o,BdG}_-({\bf K})}{2^n}\right]_{\rm floor}\notag\\
&=\sum_{{\bf K}\in {\rm TRIM}}
\left[
\frac{n^{\rm u}_+({\bf K})+n^{\rm o}_-({\bf K})}{2^n}
\right]_{\rm floor}\notag\\
&=\sum_{{\bf K}\in {\rm TRIM}}
\left[
\frac{
n^{\rm u}_+({\bf K})
+n^{\rm u}_-({\bf K})
+2n^{\rm o}_-({\bf K})
-n^{\rm o}_-({\bf K})
-n^{\rm u}_-({\bf K})
}{2^n}
\right]_{\rm floor}\notag\\
&=\sum_{{\bf K}\in {\rm TRIM}}
\left[
\frac{n^{\rm u}({\bf K})}{2^n}
+\frac{n^{\rm o}_-({\bf K})}{2^{n-1}}
-\frac{n_-({\bf K})}{2^n}
\right]_{\rm floor}\notag\\
&=\sum_{{\bf K}\in {\rm TRIM}}
\left(
\left[
\frac{n^{\rm u}({\bf K})}{2^n}
\right]_{\rm floor}
+
\left[
\frac{n^{\rm o}_-({\bf K})}{2^{n-1}}
\right]_{\rm floor}
-
\left[
\frac{n_-({\bf K})}{2^n}
\right]_{\rm floor}
+
\delta_{2^n}({\bf K})
\right)\notag\\
&=
\sum_{{\bf K}\in {\rm TRIM}}
\left[
\frac{n^{\rm u}({\bf K})}{2^n}
\right]_{\rm floor}
+
\sum_{{\bf K}\in {\rm TRIM}}
\left[
\frac{n^{\rm o}_-({\bf K})}{2^{n-1}}
\right]_{\rm floor}
+
\sum_{{\bf K}\in {\rm TRIM}}
\left[
\frac{n_-({\bf K})}{2^n}
\right]_{\rm floor}
+
\sum_{{\bf K}\in {\rm TRIM}}
\delta_{2^n}({\bf K})
\end{align}
modulo two, where $\delta_{2^n}({\bf K})$ is defined by the fourth and fifth lines, and we flip the sign of the third term in the last line, which is possible because we count only mod 2.\\

\subsection{Third-order topological superconductor in three dimensions.}
We assume time reversal $T$ and inversion $P$ symmetries that satisfy $T^2=P^2=1$ in the normal state, and an additional particle-hole $C$ symmetry that satisfies $C^2=1$ and $CP=-PC$ in the superconducting state.
Accordingly, in both normal and superconducting states, nodal points and nodal lines are stable in two and three dimensions, respectively.
In the main text, we show that, when we can take an inversion-invariant unit cell, the normal state need to obtain a weak-pairing odd-parity second-order topological superconductor with $\nu_2^{\rm BdG}=1$ in two and three dimensions is a semimetal characterized by $\nu_1=1$.
Here, we show that, when we can take a unit cell that is inversion-invariant, we need a monopole nodal line semimetal characterized by $\nu_2=1$ as a normal state to achieve a fully gapped third-order topological superconductor with $\nu_4^{\rm BdG}=1$ in three dimensions by weak odd-parity pairing.
To prove this statement, we use the following three conditions:
\begin{enumerate}
\item there is no Fermi surface enclosing a TRIM,
\item the superconducting state has no nodes requires by inversion parity, and
\item the unit cell is inversion-invariant.
\end{enumerate}
Note that the first and second conditions required to get a fully gapped superconductor.
The first condition is required to obtain a fully gapped superconducting state in the weak pairing limit, because a weak odd-parity pairing creates nodes on each Fermi surface enclosing a TRIM due to the sign change of the pairing function on the Fermi surface.
The second condition states that there is no parity-enforced node in the superconducting state.

We first write $n^{\rm u}({\bf K})$, $n^{\rm o}_-({\bf K})$, and $n_-({\bf K})$ in binary.
\begin{align}
n^{\rm u}({\bf K})
&=\alpha_1({\bf K})+2\alpha_2({\bf K})+4\alpha_4({\bf K})+ \hdots,\notag\\
n^{\rm o}_-({\bf K})
&=\beta_1({\bf K})+2\beta_2({\bf K})+4\beta_4({\bf K})+ \hdots,\notag\\
n_-({\bf K})
&=\gamma_1({\bf K})+2\gamma_2({\bf K})+4\gamma_4({\bf K})+ \hdots,
\end{align}
where $\alpha_{2^n}({\bf K})$, $\beta_{2^n}({\bf K})$, and $\gamma_{2^n}({\bf K})$ are $0$ or $1$ for all nonnegative integer $n$. Then,
\begin{align}
\nu_{2^n}
&=
\sum_{{\bf K}\in {\rm TRIM}}
\beta_{2^{n}}({\bf K}) \mod 2,
\notag\\
\nu^{\rm BdG}_{2^n}
&=
\sum_{{\bf K}\in {\rm TRIM}}
\alpha_{2^n}({\bf K})
+
\sum_{{\bf K}\in {\rm TRIM}}
\beta_{2^{n-1}}({\bf K})
+
\sum_{{\bf K}\in {\rm TRIM}}
\gamma_{2^n}({\bf K})
+
\sum_{{\bf K}\in {\rm TRIM}}
\delta_{2^n}({\bf K})
\mod 2,
\end{align}
The condition 1 and 3 requires all $\alpha_{2^n}({\bf K})$s and all $\gamma_{2^n}({\bf K})$s to be constants (independent of ${\bf K}$), respectively, i.e.,
\begin{align}
\alpha_{2^n}({\bf K})
&=\alpha_{2^n}\quad \forall n,{\bf K},\notag\\
\gamma_{2^n}({\bf K})
&=\gamma_{2^n}\quad \forall n,{\bf K}.
\end{align}
Here, $\gamma_{2^n}({\bf K})$s are constants because, in the case when we can take an inversion-invariant unit cell, all electronic states can be continuously deformed to the unit cell center, such that $n_-({\bf K})$ is independent of ${\bf K}$.
The condition 2 requires that the superconducting state does not have symmetry-required nodes, i.e., $\nu_1^{\rm BdG}=\nu_2^{\rm BdG}=0$ because $\nu_1^{\rm BdG}$ on a plane counts the number of nodal lines penetrating the plane (2D sub-Brillouin zone), and $\nu_2^{\rm BdG}$ counts the number of monopole nodal line pairs in the Brillouin zone:
\begin{align}
\label{eq:nu1nu2}
\nu_1^{\rm BdG}
&=\sum_{\substack{{\bf K}\in {\rm TRIM}\\ \rm{in\; a\; plane}}}\alpha_1({\bf K})=0 \mod 2\notag\\
\nu_2^{\rm BdG}
&=\sum_{{\bf K}\in {\rm TRIM}} \left[\alpha_2({\bf K})+\beta_1({\bf K})+\gamma_2({\bf K})+\delta_2({\bf K})\right]
=\sum_{{\bf K}\in {\rm TRIM}} \beta_1({\bf K})=0 \mod 2,
\end{align}
where, in the second line, we use $\sum_{{\bf K}\in {\rm TRIM}} \alpha_2({\bf K})=0 \mod 2$, $\sum_{{\bf K}\in {\rm TRIM}} \gamma_2({\bf K})=0 \mod 2$, and $\sum_{{\bf K}\in {\rm TRIM}} \delta_2({\bf K})=0 \mod 2$ because $\delta_2({\bf K})=[\alpha_1({\bf K})/2-\gamma_1({\bf K})/2]_{\rm floor}$ is either $\alpha_1({\bf K})$ or $1-\alpha_1({\bf K})$ depending on the value of the constant $\gamma_1({\bf K})=\gamma_1$.
While the first line is redundant as it can also be derived from the condition 1, the second line gives a new constraint on $\beta_1({\bf K})$.

It immediately follows that
\begin{align}
\sum_{{\bf K}\in {\rm TRIM}}
\alpha_{4}({\bf K})
&
=\alpha_{4}
\sum_{{\bf K}\in {\rm TRIM}} 1
=0 \mod 2,\notag\\
\sum_{{\bf K}\in {\rm TRIM}}
\gamma_{4}({\bf K})
&
=\gamma_{4}
\sum_{{\bf K}\in {\rm TRIM}} 1
=0 \mod 2.
\end{align}
Also, we have
\begin{align}
\label{eq:delta=0}
\sum_{{\bf K}\in {\rm TRIM}}\delta_4({\bf K})
=\sum_{{\bf K}\in {\rm TRIM}}\left[\frac{1}{4}\alpha_1+\frac{1}{2}\alpha_2-\frac{1}{4}\gamma_1-\frac{1}{2}\gamma_2+\frac{1}{2}\beta_1({\bf K})\right]_{\rm floor}=0 \mod 2.
\end{align}
This can be shown as follows.
First, note that $\delta_4({\bf K})=\left[\alpha_1/4+\alpha_2/2-\gamma_1/4-\gamma_2/2+\beta_1({\bf K})/2\right]_{\rm floor}$ is either independent of $\beta_1({\bf K})$ or dependent on the value of $\beta_1({\bf K})$. 
In the former, since $\delta_4({\bf K})$ is an integer that is independent of ${\bf K}$, Eq.~\eqref{eq:delta=0} is obviously valid.
In the latter, $\delta_4({\bf K})$ can be either $\beta_1({\bf K}) \mod 2$ or $1-\beta_1({\bf K}) \mod 2$, so $\sum_{\bf K}\delta_4({\bf K})=0 \mod 2$ is guaranteed by the second line in Eq.~\eqref{eq:nu1nu2}.
Thus, we have Eq.~\eqref{eq:delta=0}.

In conclusion, $\nu_4^{\rm BdG}=1$ only when the normal state is a monopole nodal line semimetal characterized by $\nu_2=1$ when the three conditions stated in the beginning of this subsection are satisfied, because then
\begin{align}
\nu^{\rm BdG}_{4}
&=
\sum_{{\bf K}\in {\rm TRIM}}
\alpha_{4}({\bf K})
+
\sum_{{\bf K}\in {\rm TRIM}}
\beta_{2}({\bf K})
+
\sum_{{\bf K}\in {\rm TRIM}}
\gamma_{4}({\bf K})
+
\sum_{{\bf K}\in {\rm TRIM}}
\delta_{4}({\bf K})
\mod 2\notag\\
&=
\sum_{{\bf K}\in {\rm TRIM}}
\beta_{2}({\bf K})
\mod 2\notag\\
&=
\sum_{{\bf K}\in {\rm TRIM}}
\left[
\frac{n^{\rm o}_-({\bf K})}{2}
\right]_{\rm floor} 
\mod 2\notag\\
&=\nu_2\mod 2.
\end{align}

\section{General Form of Model Hamiltonians}

Here we write down the most general form of the two-, three-, and four-band normal state Hamiltonians with inversion $P$ and time reversal $T$ symmetries with $P^2=T^2$ and their odd-parity BdG Hamiltonians whose particle-hole operator $C$ satisfies $C^2=1$.
The two- and four-band normal state models are used in In Sec.~\ref{sec:bulk-boundary}, and the three-band normal state Hamiltonian is studied here for a discussion on the tight-binding model on the Kagome lattice in Sec.~\ref{sec:2DTB}.
In our notations, $\sigma_{i=x,y,z}$ and $\rho_{i=x,y,z}$ are Pauli matrices for the orbital degrees of freedom, $\lambda_{i=0,\hdots 8}$ are Gell-Mann matrices for the orbital degrees of freedom, and $\tau_{i=x,y,z}$ are Pauli matrices for the Nambu space.

\subsection{Two-band normal state and odd-parity pairing}

We take inversion $P$ and time reversal $T$ symmetry operators as
\begin{align}
P=\sigma_z,\quad
T=K.
\end{align}
Then, the $2\times 2$ normal state Hamiltonian has the form of
\begin{align}
h({\bf k})
=-\mu({\bf k})+f_1({\bf k})\sigma_y+f_2({\bf k})\sigma_z,
\end{align}
where
\begin{align}
\mu({\bf k})
&=+\mu(-{\bf k})\notag\\
f_1({\bf k})
&=-f_1(-{\bf k}),\notag\\
f_2({\bf k})
&=+f_2(-{\bf k}).
\end{align}
The energy spectrum is given by
\begin{align}
E({\bf k})
&=-\mu({\bf k})\pm \sqrt{f_1^2({\bf k})+f_2^2({\bf k})}.
\end{align}

In the case of odd-parity pairing, inversion operator acts on the particle and hole sector with a different sign, so particle-hole $C$, inversion $P$, and time reversal $T$ symmetry operators are
\begin{align}
C
=\tau_xK,\quad
P
=\tau_z\sigma_z,\quad
T
=K.
\end{align}
The $4\times 4$ BdG Hamiltonian compatible with those symmetries is
\begin{align}
\label{eq:4BdG}
H({\bf k})
&=
-\mu({\bf k})\tau_z
+f_1({\bf k})\tau_z\sigma_y
+f_2({\bf k})\tau_z\sigma_z
+\Delta_1({\bf k})\tau_y
+\Delta_2({\bf k})\tau_y\sigma_y
+\Delta_3({\bf k})\tau_y\sigma_z
\notag\\
&=
f_1({\bf k}) \Gamma_{1}
+f_2({\bf k})\Gamma_{2}
+\Delta_1({\bf k})\Gamma_{3}
+\Delta_2({\bf k})\Gamma_{14}
+\Delta_3({\bf k})\Gamma_{24}
+\mu({\bf k})\Gamma_{34},
\end{align}
where
\begin{align}
\Delta_{i=2}({\bf k})
&=+\Delta_{i=2}(-{\bf k}),\notag\\
\Delta_{i=1,3}({\bf k})
&=-\Delta_{i=1,3}(-{\bf k}),
\end{align}
and we define mutually anticommuting Gamma matrices by
\begin{align}
\Gamma_1
&=\tau_z\sigma_y,\notag\\
\Gamma_2
&=\tau_z\sigma_z,\notag\\
\Gamma_3
&=\tau_y,\notag\\
\Gamma_4
&=\tau_x,\notag\\
\Gamma_5
&=\tau_z\sigma_x,
\end{align}
and $\Gamma_{ij}=-i\Gamma_i\Gamma_j$.
The spectrum of the BdG Hamiltonian is given by~\cite{bzdusek2017robust}
\begin{align}
E({\bf k})
&=\pm \sqrt{{\bf a}^2+{\bf b}^2\pm 2|{\bf a}\times {\bf b}|},\notag\\
\end{align}
where
\begin{align}
{\bf a}
&=(f_1,f_2,\Delta_1,),\notag\\
{\bf b}
&=(\Delta_2,\Delta_3,\mu).
\end{align}

\subsection{Four-band normal state and odd-parity pairing}

As above, we take symmetry operators as
\begin{align}
P
=\sigma_z,\quad
T
=K.
\end{align}
Then, the $4\times 4$ normal state Hamiltonian has the form of
\begin{align}
\label{eq:4N}
h({\bf k})
&=-\mu({\bf k})
+f_1({\bf k})\rho_y\sigma_x
+f_2({\bf k})\sigma_y
+f_3({\bf k})\sigma_z\notag\\
&\quad
+m_1({\bf k})\rho_z
+m_2({\bf k})\rho_x
+m_3({\bf k})\rho_x\sigma_z
+m_4({\bf k})\rho_z\sigma_z
+m_5({\bf k})\rho_x\sigma_y
+m_6({\bf k})\rho_z\sigma_y
\notag\\
&=-\mu({\bf k})
+f_1({\bf k})\gamma_1
+f_2({\bf k})\gamma_2
+f_3({\bf k})\gamma_3\notag\\
&\quad
-m_1({\bf k})\gamma_{14}
+m_2({\bf k})\gamma_{15}
-m_3({\bf k})\gamma_{24}
-m_4({\bf k})\gamma_{25}
+m_5({\bf k})\gamma_{34}
+m_6({\bf k})\gamma_{35},
\end{align}
where 
\begin{align}
\mu({\bf k})
&=+\mu(-{\bf k}),\notag\\
f_{i=1,2}({\bf k})
&=-f_{i=1,2}(-{\bf k}),\notag\\
f_3({\bf k})
&=+f_3(-{\bf k}),\notag\\
m_{i=1,2,3,4}({\bf k})
&=+m_{i=1,2,3,4}(-{\bf k}),\notag\\
m_{i=5,6}({\bf k})
&=-m_{i=5,6}(-{\bf k}),
\end{align}
and we define
\begin{align}
\gamma_1
&=\rho_y\sigma_x,\notag\\
\gamma_2
&=\sigma_y,\notag\\
\gamma_3
&=\sigma_z,\notag\\
\gamma_4
&=\rho_x\sigma_x,\notag\\
\gamma_5
&=\rho_z\sigma_x.
\end{align}

For odd-parity pairing,
\begin{align}
C=\tau_xK,\quad
P=\tau_z\sigma_z,\quad
T=K,
\end{align}
the $8\times 8$ BdG Hamiltonian is
\begin{align}
\label{eq:8BdG}
H({\bf k})
&=
-\mu({\bf k})\tau_z
+f_1({\bf k})\tau_z\rho_y\sigma_x
+f_2({\bf k})\tau_z\sigma_y
+f_3({\bf k})\tau_z\sigma_z\notag\\
&\quad
+m_1({\bf k})\tau_z\rho_z
+m_2({\bf k})\tau_z\rho_x
+m_3({\bf k})\tau_z\rho_x\sigma_z
+m_4({\bf k})\tau_z\rho_z\sigma_z
+m_5({\bf k})\tau_z\rho_x\sigma_y
+m_6({\bf k})\tau_z\rho_z\sigma_y\notag\\
&\quad
+\Delta_1({\bf k})\tau_y
+\Delta_2({\bf k})\tau_y\rho_y\sigma_x
+\Delta_3({\bf k})\tau_y\sigma_y
+\Delta_4({\bf k})\tau_y\sigma_z\notag\\
&\quad
+\Delta_5({\bf k})\tau_y\rho_z
+\Delta_6({\bf k})\tau_y\rho_x
+\Delta_7({\bf k})\tau_y\rho_x\sigma_z
+\Delta_8({\bf k})\tau_y\rho_z\sigma_z
+\Delta_9({\bf k})\tau_y\rho_x\sigma_y
+\Delta_{10}({\bf k})\tau_y\rho_z\sigma_y\notag\\
&=
f_1({\bf k}) \Gamma_{1}
+f_2({\bf k})\Gamma_{2}
+f_3({\bf k})\Gamma_{3}
+\Delta_1({\bf k})\Gamma_{4}\notag\\
&\quad
+m_1({\bf k})\Gamma_{237}
+m_2({\bf k})\Gamma_{236}
-m_3({\bf k})\Gamma_{137}
+m_4({\bf k})\Gamma_{136}
-m_5({\bf k})\Gamma_{127}
+m_6({\bf k})\Gamma_{126}\notag\\
&\quad
+\Delta_2({\bf k})\Gamma_{15}
+\Delta_3({\bf k})\Gamma_{25}
+\Delta_4({\bf k})\Gamma_{35}
+\mu({\bf k})\Gamma_{45}\notag\\
&\quad
+\Delta_5({\bf k})\Gamma_{146}
-\Delta_6({\bf k})\Gamma_{147}
+\Delta_7({\bf k})\Gamma_{246}
+\Delta_8({\bf k})\Gamma_{247}
-\Delta_9({\bf k})\Gamma_{346}
-\Delta_{10}({\bf k})\Gamma_{347},
\end{align}
where
\begin{align}
\Delta_{i=1,4,5,6,7,8}({\bf k})
&=-\Delta_{i=1,4,5,6,7,8}(-{\bf k}),\notag\\
\Delta_{i=2,3,9,10}({\bf k})
&=+\Delta_{i=2,3,9,10}(-{\bf k}),
\end{align}
and we define
\begin{align}
\Gamma_1
&=\tau_z\rho_y\sigma_x,\notag\\
\Gamma_2
&=\tau_z\sigma_y,\notag\\
\Gamma_3
&=\tau_z\sigma_z,\notag\\
\Gamma_4
&=\tau_y,\notag\\
\Gamma_5
&=\tau_x,\notag\\
\Gamma_6
&=\tau_z\rho_x\sigma_x,\notag\\
\Gamma_7
&=\tau_z\rho_z\sigma_x.\notag\\
\end{align}

\subsection{Three-band normal state and odd-parity pairing}

We consider the following represetations of $P$ and $T$
\begin{align}
P
=1,\quad
T
=K,
\end{align}
which is relevant for the Kagome lattice we study below.
The most general form of the $3\times 3$ normal state Hamiltonian is then
\begin{align}
\label{eq:3N}
h({\bf k})
=
-\mu_1({\bf k})\lambda_0
-\mu_2({\bf k})\lambda_3
-\mu_3({\bf k})\lambda_8
+f_1({\bf k})\lambda_1
+f_2({\bf k})\lambda_4
+f_3({\bf k})\lambda_6,
\end{align}
where
\begin{align}
\mu_{i=1,2,3}({\bf k})
&=+\mu_{i=1,2,3}(-{\bf k}),\notag\\
f_{i=1,2,3}({\bf k})
&=+f_{i=1,2}(-{\bf k}).
\end{align}

For odd-parity pairing, symmetry operators are
\begin{align}
C
=\tau_xK,\quad
P
=\tau_z,\quad
T
=K,
\end{align}
and the $6\times 6$ BdG Hamiltonian symmetric under those operations takes the form of
\begin{align}
\label{eq:6BdG}
H({\bf k})
&=
-\mu_1({\bf k})\tau_z\lambda_0
-\mu_2({\bf k})\tau_z\lambda_3
-\mu_3({\bf k})\tau_z\lambda_8
+f_1({\bf k})\tau_z\lambda_1
+f_2({\bf k})\tau_z\lambda_4
+f_3({\bf k})\tau_z\lambda_6\notag\\
&\quad 
+\Delta_1({\bf k})\tau_y\lambda_0
+\Delta_2({\bf k})\tau_y\lambda_3
+\Delta_3({\bf k})\tau_y\lambda_8
+\Delta_4({\bf k})\tau_y\lambda_1
+\Delta_5({\bf k})\tau_y\lambda_4
+\Delta_6({\bf k})\tau_y\lambda_6,
\end{align}
where
\begin{align}
\Delta_{i=1,...,6}({\bf k})
=-\Delta_{i=1,...,6}(-{\bf k}).
\end{align}

\section{Tight-Binding Models}
\label{sec:bulk-boundary}

\begin{figure}[b!]
\includegraphics[width=16cm]{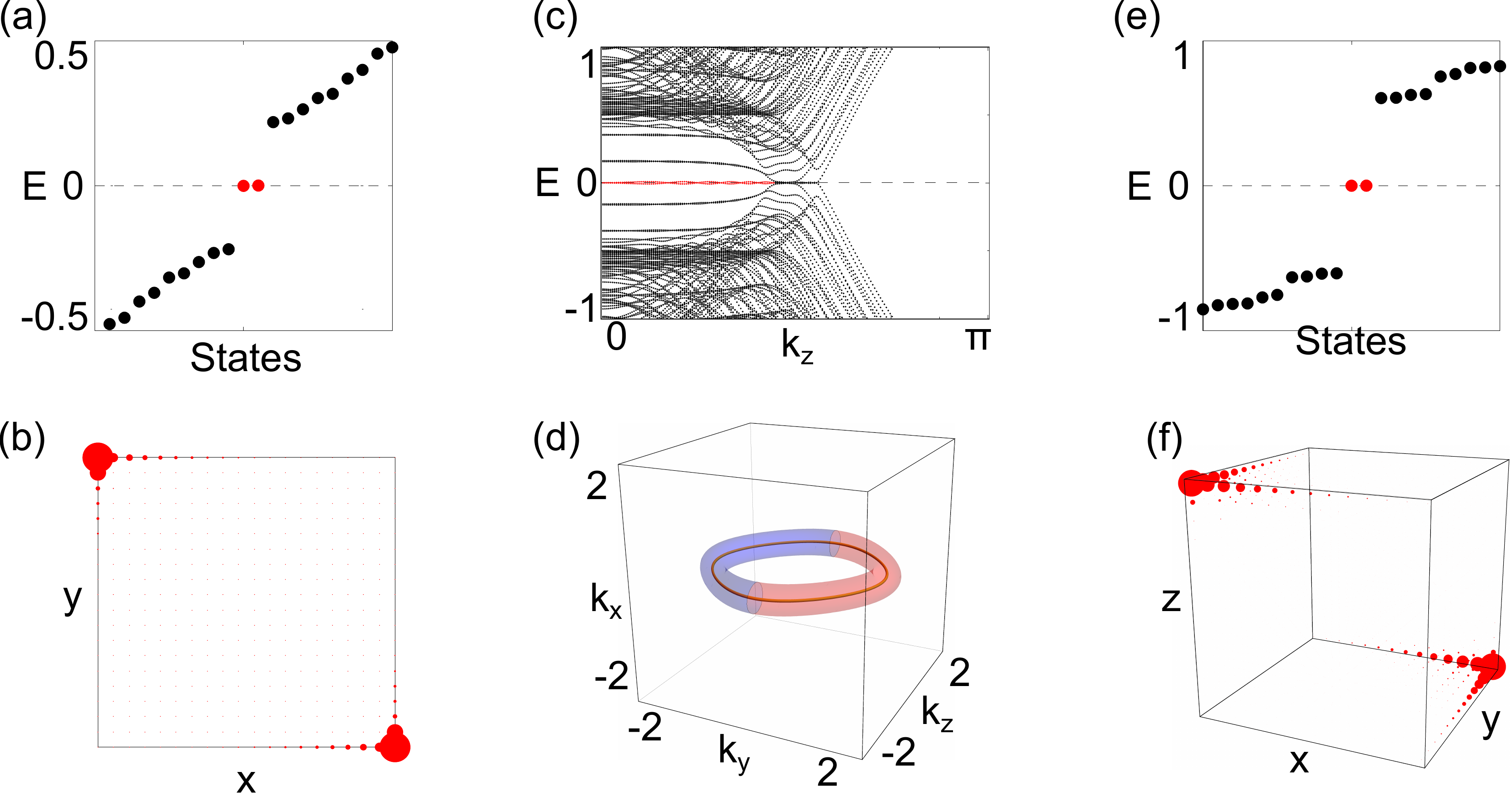}
\caption{Higher-order topological superconductivity.
(a,b) Zero modes and boundary Majorana states of a two-dimensional second-order topological superconductor described in Eq.~\eqref{eq:TB_2D}.
Here $t=1$, $\mu=0.2$, $M=2$, and $\Delta_{1,3}({\bf k})=\sum_{i=x,y}(\Delta^V_{i,ss}\pm \Delta^V_{i,p_xp_x})\sin k_i$, and $\Delta_2({\bf k})=\Delta^U$, where $\Delta^V_{x,ss}=\Delta^V_{x,p_xp_x}=0$, $\Delta^V_{y,ss}=\Delta^V_{y,p_xp_x}=0.5$, and $\Delta^U=0.5$, and a $20\times 20$ lattice is considered.
(c,d) Zero modes and nodal lines of a monopole nodal line superconductor described in Eq.~\eqref{eq:TB_2D}.
We take $t=1$, $M=4$, $\mu=0.2$, and 
$\Delta_{1,3}({\bf k})=\sum_{i=x,y,z}(\Delta^V_{i,ss}\pm \Delta^V_{i,p_xp_x})\sin k_i$, and $\Delta_2({\bf k})=\Delta^U$, where $\Delta^V_{x,ss}=\Delta^V_{x,p_xp_x}=\Delta^V_{z,ss}=\Delta^V_{z,p_xp_x}=0$, $\Delta^V_{y,ss}=\Delta^V_{y,p_xp_x}=0.2$, and $\Delta^U=0$, and a lattice with $20\times 20$ unit cells along $x$ and $y$ in (c).
In (d), the torus indicates the Fermi surface enclosing a nodal line (thick gold line) in the normal state. The blue (red) color indicates the region where the pairing function has positive (negative) sign. Two monopole nodal loops appear at the intersection, where the pairing function changes its sign.
(e,f) Zero modes and boundary Majorana states of a three-dimensional third-order topological superconductor described in Eq.~\eqref{eq:TB_3D3rd}. 
We take $t=1$, $\mu=0.2$, $M=4$, $m_1=m_2=\Delta_1=\Delta_2=\Delta_3=0.5$ on the $14\times 14\times 14$ lattice.
}
\label{fig:higher-order}
\end{figure}

We demonstrate our theory by using simple tight-binding models defined on rectangular or orthorhombic lattices.
We construct three models in which the normal states are a Dirac semimetal in 2D, a nodal line semimetal in 3D, and a monopole nodal line semimetal in 3D, respectively.
We show that, when an odd-parity superconducting pairing is introduced, the three nodal semimetals turn into a second-order topological superconductor in 2D, a monopole nodal line superconductor in 3D, and a third-order topological superconductor in 3D, respectively.

\subsection{Dirac semimetal and second-order topological superconductor in two dimensions}

First, let us consider a 2D Dirac semimetal described by the nearest-neighbor tight-binding Hamiltonian
\begin{align}
\hat{h}
=\hat{h}_{\mu}+\hat{h}_t
=
-\sum_{i,\sigma}\mu_{\sigma}c^{\dagger}_{i,\sigma}c_{i,\sigma}
-\sum_{\braket{ij};\sigma,\sigma'}t_{i,\sigma;j,\sigma'}c^{\dagger}_{i,\sigma}c_{j,\sigma'},
\end{align}
where $\sigma=s,p_x$ labels the orbital degrees of freedom.
In momentum space, we have
\begin{align}
h
&=
-\mu+2t\sin k_x\sigma_y+(M-2t\cos k_x-2t\cos k_y)\sigma_z,
\end{align}
where $\sigma_{y,z}$ are the Pauli matrices for orbital degrees of freedom with $\uparrow$ ($\downarrow$) indicating a $s$ ($p_x$) orbital. 
\begin{align}
\mu
&=\frac{1}{2}\left(\mu_s+\mu_{p_x}\right),\notag\\
M
&=-\frac{1}{2}\left(\mu_s-\mu_{p_{x}}\right)\notag\\
t
&=t_{ss}=-t_{p_xp_x}=t_{sp_x}.
\end{align}
The Hamiltonian is symmetric under $P=\sigma_z$, $T=K$, $M_x=\sigma_z$, $M_y=1$, $M_z=1$, where $M_{i=x,y,z}$ indicates a mirror operator that flips the sign of the $i$-th coordinate.
The energy eigenvalues are
$
E({\bf k})
=-\mu\pm\sqrt{4t^2\sin^2 k_x+(M-2t\cos k_x-2t\cos k_y)^2}
$
which exhibit two Dirac points at $(k_x,k_y)=(0,\pm \cos^{-1}[(M/t-2)/2])$ when $0<M/t<4$ at the energy $E=-\mu$.

To induce superconductivity, we consider the following interaction term 
\begin{align}
H_{\rm int}
&=
-U\sum_{i,\sigma,\sigma'}n_{i,\sigma}n_{i,\sigma'}
-V\sum_{\braket{i,j},\sigma}n_{i,\sigma}n_{j,\sigma},
\end{align}
where $U$ indicates the on-site interorbital interaction and $V$ denotes the intraorbital interaction between 
nearest-neighbor sites. 
We treat $H_{\rm int}$ within the mean field approximation.
Although we focus only inversion symmetry, for a systematic analysis of the lattice model, let us first organize
the odd-parity pairing in terms of the irreducible representations of the $D_{2h}$ symmetry group as
\begin{align}
B_{2u}: 
&-i\left(\Delta^V_{y,ss}+\Delta^V_{y,p_xp_x}\right)\sin k_y
-i\left(\Delta^V_{y,ss}-\Delta^V_{y,p_xp_x}\right)\sin k_y\sigma_z,
\notag\\
B_{3u}:
&-i\left(\Delta^V_{x,ss}+\Delta^V_{x,p_xp_x}\right)\sin k_x
-i\left(\Delta^V_{x,ss}-\Delta^V_{x,p_xp_x}\right)\sin k_x\sigma_z+i\Delta^U\sigma_y,
\end{align}
where $B_{2u}$ has parity $(+,-,+)$ under $(M_x,M_y,M_z)$, and $B_{3u}$ has parity $(-,+,+)$ under $(M_x,M_y,M_z)$.
The $B_{2u}$ pairing fully opens the gap.
On the other hand, the $B_{3u}$ pairing generates four nodes because the intersection of Fermi surfaces and the $k_x=0$ plane are the crossing points between two bands with different $M_x$ eigenvalues in the BdG spectrum (by the same argument applied to odd-parity pairing, the particle sector and the hole sector in the Nambu space have opposite $M_x$ eigenvalues since $B_{3u}$ is an odd-$M_x$ pairing).
Accordingly, we need to include nonvanishing $B_{2u}$ pairing to open the gap.
In general the BdG Hamiltonian takes the form
\begin{align}
\label{eq:TB_2D}
H_{\rm BdG}({\bf k})
=
&-\mu\tau_z+2t\sin k_x\tau_z\sigma_y
+(M-2t\cos k_x-2t\cos k_y)\tau_z\sigma_z\notag\\
&
+\Delta_1({\bf k})\sin k_y\tau_y
+\Delta_2({\bf k})\tau_y\sigma_y
+\Delta_{3}({\bf k}) \sin k_y\tau_y\sigma_z,
\end{align}
where $\tau_{i=x,y,z}$ are the Pauli matrices for the Nambu space, $\Delta_{1,3}({\bf k})=\sum_{i=x,y}(\Delta^V_{i,ss}\pm \Delta^V_{i,p_xp_x})\sin k_i$, and $\Delta_2({\bf k})=\Delta^U$.
We take $t=1$, $\mu=0.2$, $M=2$, $\Delta^V_{x,ss}=\Delta^V_{x,p_xp_x}=0$, $\Delta^V_{y,ss}=\Delta^V_{y,p_xp_x}=0.5$, and $\Delta^U=0.5$, and consider a $20\times 20$ lattice for numerical calculations.
The results in Fig.~\ref{fig:higher-order}(a,b) show two zero modes localized at two corners.

\subsection{Nodal line semimetal and second-order topological nodal superconductor in three dimensions}

Similalry, we can construct the Hamiltonian for a 3D nodal line semimetal 
as
\begin{align}
h({\bf k})
=
&-\mu+2t\sin k_x\sigma_y
+(M-2t\cos k_x-2t\cos k_y-2t\cos k_z)\sigma_z,
\end{align}
which can be considered as a vertical stacking of 2D Dirac semimetals with an additonal hopping along the $z$ direction.
The energy eigenvalues are $E({\bf k})
=-\mu\pm\sqrt{4t^2\sin^2 k_x+M_{\bf k}^2}$, where $M_{\bf k}=M-2t\cos k_x-2t\cos k_y-2t\cos k_z$.
A single nodal loop surrounding the $\Gamma$ point appears in the $k_x=0$ plane at the energy $E=-\mu$ when $2<M/t<6$.
When $|\mu|$ is smaller than the bandwidth, the Fermi surface is torus-shaped.
If we include on-site interorbital $U$ and nearest-neighbor intraorbital $V$ Coulomb interactions as in two dimensions, odd-parity pairing terms are organized into the $D_{2h}$ irreducible representations as
\begin{align}
B_{1u}: 
&-i\left(\Delta^V_{z,ss}+\Delta^V_{z,p_xp_x}\right)\sin k_z
-i\left(\Delta^V_{z,ss}-\Delta^V_{z,p_xp_x}\right)\sin k_z\sigma_z,
\notag\\
B_{2u}: 
&-i\left(\Delta^V_{y,ss}+\Delta^V_{y,p_xp_x}\right)\sin k_y
-i\left(\Delta^V_{y,ss}-\Delta^V_{y,p_xp_x}\right)\sin k_y\sigma_z,
\notag\\
B_{3u}:
&-i\left(\Delta^V_{x,ss}+\Delta^V_{x,p_xp_x}\right)\sin k_x
-i\left(\Delta^V_{x,ss}-\Delta^V_{x,p_xp_x}\right)\sin k_x\sigma_z+i\Delta^U\sigma_y,
\end{align}
where $B_{1u}$,  $B_{2u}$, and $B_{3u}$ has parity $(+,+,-)$, $(+,-,+)$, and $(-,+,+)$ under $(M_x,M_y,M_z)$, respectively.
The relevant BdG Hamiltonian is given by
\begin{align}
\label{eq:TB_3DMNL}
H_{\rm BdG}({\bf k})
=
&-\mu\tau_z+2t\sin k_x\tau_z\sigma_y
+(M-2t\cos k_x-2t\cos k_y-2t\cos k_z)\tau_z\sigma_z\notag\\
&+\Delta_1({\bf k})\sin k_y\tau_y
+\Delta_2({\bf k})\tau_y\sigma_y
+\Delta_{3}({\bf k}) \sin k_y\tau_y\sigma_z,
\end{align}
 with $\Delta_{1,3}({\bf k})=\sum_{i=x,y,z}(\Delta^V_{i,ss}\pm \Delta^V_{i,p_xp_x})\sin k_i$ and $\Delta_2({\bf k})=\Delta^U$.
We take $t=1$, $M=4$, $\mu=0.2$ with a $B_{2u}$ pairing: $\Delta^V_{x,ss}=\Delta^V_{x,p_xp_x}=\Delta^V_{z,ss}=\Delta^V_{z,p_xp_x}=0$, $\Delta^V_{y,ss}=\Delta^V_{y,p_xp_x}=0.2$, and $\Delta^U=0$ for numerical calculations.
The spectrum of the BdG Hamiltonian is gapless at $k_y=0$ on the torus Fermi surface due to the sign change of the pairing term.
The nodal structure can be seen explicitly from the energy eigenvalues of the BdG Hamiltonian given by 
$
\xi({\bf k})
=\pm \sqrt{\left(\Delta_{y,ss}^V+\Delta_{y,p_xp_x}^V\right)^2\sin^2k_y +\left(\sqrt{4t^2\sin^2 k_x+M_{\bf k}^2}-\mu\right)^2}.
$
Fig.~\ref{fig:higher-order}(d) shows the corresponding Fermi surface of a torus shape and the location of nodes.
Interestingly, the two nodal loops at $k_y=0$ are linked with the nodal loop of the normal state, which appears as the crossing between the occupied bands of the BdG Hamiltonian. This \noindent{\it linking structure} indicates that the zero-energy nodes carry nontrivial monopole charges~\cite{ahn2018band}.
To see the higher-order topology of this phase, we consider the open boundary condition with $20\times 20$ unit cells along $x$ and $y$ and the periodic boundary condition along the $z$ direction.
The spectrum shown in Fig.~\ref{fig:higher-order}(c) reveals zero modes in a finite range of $k_z$, which is inside the nodal loop of the normal state, and they originate from the nontrivial second Stiefel-Whitney number in the range of $k_z$.
We also have two monopole nodal lines for pairing in the $B_{1u}$ representation, which are now on the $k_z=0$ plane.
Let us note that, for pairing in the $B_{3u}$ representation, however, two trivial loops are created at the $k_x=0$ plane.
Although $\nu_2^{\rm BdG}=1$, monopole nodal line do not exist.
It is basically because any inversion-invariant 2D subBrillouin zone passing through the ${\bf k}={\bf 0}$ is gaplesss.

\subsection{Monopole nodal line semimetal and third-order topological superconductor in three dimensions}

Finally, let us consider the odd-parity superconductivity of a doped monopole nodal line semimetal. The nearest-tight-binding Hamiltonian for the normal state is
\begin{align}
\hat{h}
=\hat{h}_{\mu}+\hat{h}_t
=
-\sum_{i,\tau}\mu_{\tau}c^{\dagger}_{i,\tau}c_{i,\tau}
-\sum_{\braket{ij};\tau,\tau'}t_{i,\tau;j,\tau'}c^{\dagger}_{i,\tau}c_{j,\tau'},
\end{align}
where $\tau=s,p_x,p_y,d_{xy}$ labels orbital degrees of freedom.
In momentum space,
\begin{align}
h({\bf k})
=
&-\mu+2t\sin k_x\rho_y\sigma_x+2t\sin k_y\sigma_y
+(M-2t\cos k_x-2t\cos k_y-2t\cos k_z)\sigma_z\notag\\
&+m_1\rho_z+m_2\rho_z\sigma_z
\end{align}
where $\rho$ and $\sigma$ are Pauli matrices for orbital degrees of freedom.
Here, the basis states are $(s,p_y,d_{xy},p_x)^T$, and
\begin{align}
\mu
&=\frac{1}{4}\left(\mu_s+\mu_{p_y}+\mu_{d_{xy}}+\mu_{p_x}\right),\notag\\
M
&=-\frac{1}{4}\left(\mu_s-\mu_{p_y}+\mu_{d_{xy}}-\mu_{p_x}\right),\notag\\
m_1
&=-\frac{1}{4}\left(\mu_s+\mu_{p_y}-\mu_{d_{xy}}-\mu_{p_x}\right),\notag\\ 
m_2
&=-\frac{1}{4}\left(\mu_s-\mu_{p_y}-\mu_{d_{xy}}+\mu_{p_x}\right),\notag\\
t
&=t_{ss}=-t_{p_yp_y}=t_{d_{xy}d_{xy}}=-t_{p_xp_x}=t_{sp_x}=t_{d_{xy}p_y}=t_{sp_y}=t_{d_{xy}p_x}.
\end{align}
The Hamiltonian is symmetric under $P=\sigma_z$, $T=K$, $M_x=\rho_z$, $M_y=\rho_z\sigma_z$, $M_z=1$.
When $m_1=m_2=0$ and $2<M/t<6$, this Hamiltonian describes two Dirac points at $(k_x,k_y,k_z)=(0,0,\pm \cos^{-1}[(M-4)/2t])$, which correspond to a particular limit of a monopole nodal line shrunken to a point.
It can bee seen from the energy spectrum $E({\bf k})=-\mu\pm\sqrt{4t^2\sin^2 k_x+4t^2\sin^2 k_y+M_{\bf k}^2}$, where $M_{\bf k}=M-2t\cos k_x-2t\cos k_y-2t\cos k_z$.
Nonzero $m_1$ and $m_2$ make the monopole nodal lines have a finite size.
When interorbital onsite and intraorbital nearest-neighbor Coulomb interactions are considered, pairing functions can be classified to $B_{1u}$, $B_{2u}$, or $B_{3u}$ representation of $D_{2h}$ as we did above, and the $B_{1u}$ pairing opens the full gap.
Therefore we consider the $B_{1u}$ pairing with two additional mirror-breaking pairing $\Delta_9$ and $\Delta_{10}$ needed to obtain well-localized corner states.
Adding those superconducting pairing, we have
\begin{align}
\label{eq:TB_3D3rd}
H_{\rm BdG}({\bf k})
=
&
-\mu\Gamma_{45}
+2t\sin k_x\Gamma_1
+2t\sin k_y\Gamma_2
+(M-2t\cos k_x-2t\cos k_y-2t\cos k_z)\Gamma_3\notag\\
&+m_{1}\Gamma_{136}
+m_{2}\Gamma_{237}
+\Delta_1 \sin k_z\Gamma_4
+\Delta_{4}\sin k_z \Gamma_{35}\notag\\
&
+\Delta_{5}\sin k_z\Gamma_{146}
+\Delta_{8}\sin k_z\Gamma_{247}
-\Delta_{9}\Gamma_{346}
-\Delta_{10}\Gamma_{347},
\end{align}
where
$
\Gamma_1
=\tau_z\rho_y\sigma_x,
\Gamma_2
=\tau_z\sigma_y,
\Gamma_3
=\tau_z\sigma_z,
\Gamma_4
=\tau_y,
\Gamma_5
=\tau_x,
\Gamma_6
=\tau_z\rho_x\sigma_x,
\Gamma_7
=\tau_z\rho_z\sigma_x,
$
$\Gamma_{ij}=-i\Gamma_i\Gamma_j$, and $\Gamma_{ijk}=-i\Gamma_i\Gamma_j\Gamma_k$.
The spectrum is fully gapped
when $t=1$, $\mu=0.2$, $M=4$, $m_1=m_2=\Delta_1=\Delta_9=\Delta_{10}=0.5$, and $\Delta_4=\Delta_5=\Delta_8=0$.
Fig.~\ref{fig:higher-order}(e,f) shows the presence of zero modes localized at two corners.

\section{More on Tight-Binding Models in Two Dimensions}
\label{sec:2DTB}

\subsection{Hexagonal lattice}

\subsubsection{Normal state}

We begin with the nearest-neighbor tight-binding model.
\begin{align}
\hat{h}
=\hat{h}_{\mu}+\hat{h}_t
=
-\mu\sum_{i}c^{\dagger}_{i}c_{i}
-t\sum_{\braket{ij}}c^{\dagger}_{i}c_{j}
\end{align}
In momentum space, we have
\begin{align}
h({\bf k})
&=
\begin{pmatrix}
\mu	& -t\left(e^{-i{\bf k}\cdot{\bf a}_1}+e^{-i{\bf k}\cdot{\bf a}_2}+e^{-i{\bf k}\cdot{\bf a}_3}\right)\\
-t\left(e^{i{\bf k}\cdot{\bf a}_1}+e^{i{\bf k}\cdot{\bf a}_2}+e^{i{\bf k}\cdot{\bf a}_3}\right)	&\mu
\end{pmatrix}
\notag\\
&=
\mu
-[t\cos({\bf k}\cdot{\bf a}_1)
+t\cos({\bf k}\cdot{\bf a}_2)
+t\cos({\bf k}\cdot{\bf a}_3)]\sigma_x
-[t\sin({\bf k}\cdot{\bf a}_1)
+t\sin({\bf k}\cdot{\bf a}_2)
+t\sin({\bf k}\cdot{\bf a}_3)]\sigma_y\notag\\
&=
\mu
-\left(t\cos k_1
+t\cos k_2
+t\cos k_3\right)\sigma_x
-\left(t\sin k_1
+t\sin k_2
+t\sin k_3\right)\sigma_y
\end{align}
where
\begin{align}
k_1
&
={\bf k}\cdot{\bf a_1}
=\frac{1}{\sqrt{3}}k_y,\notag\\
k_2
&
={\bf k}\cdot{\bf a_2}
=-\frac{1}{2}k_x-\frac{1}{2\sqrt{3}}k_y\notag\\
k_3
&
={\bf k}\cdot{\bf a_3}
=\frac{1}{2}k_x-\frac{1}{2\sqrt{3}}k_y.
\end{align}
It has symmetries under
\begin{align}
T
=K,\quad
C_3
=1,\quad
M_z
=1,\quad
M_x
=1,\quad
M_y
=\sigma_x,\quad
P
=\sigma_x,
\end{align}
where the crystalline symmetry operations form the $D_{6h}$ group.
The Hamiltonian has the form of
\begin{align}
h({\bf k})
=-\mu({\bf k})+f_1({\bf k})\sigma_x+f_2({\bf k})\sigma_y,
\end{align}
where
\begin{align}
\mu({\bf k})
&=\mu,\notag\\
f_1({\bf k})
&=
-t\cos k_1-t\cos k_2-t\cos k_3,\notag\\
f_2({\bf k})
&=
-t\sin k_1-t\sin k_2-t\sin k_3.
\end{align}
Two Dirac points appear at $K=(\pi/\sqrt{3},\pi)$ and $K'=(-\pi/\sqrt{3},\pi)$ points at $E=-\mu$.

\begin{align}
C
=\tau_xK,\quad
P
=\tau_z\sigma_x.
\end{align}

\subsubsection{Superconducting state}

Let us consider Coulomb interactions.
There is no on-site Coulomb interaction because of the Fermi statistics.
We consider the nearest and the next-nearest neighbor Coulomb interactions.

\begin{align}
H_{\rm int}
=&
-U\sum_{\braket{i,j}}n_in_j
-V\sum_{\braket{\braket{i,j}}}n_in_j\notag\\
\approx&
\sum_{\braket{i,j}}\bigg[
c^{\dagger}_{i}c^{\dagger}_{j}\Delta^U_{ij}
+c_{j}c_{i}\Delta^{U*}_{ij}
-U^{-1}\Delta^{U*}_{ij}\Delta^U_{ij}
\bigg]
+
\sum_{\braket{\braket{i,j}}}\bigg[
c^{\dagger}_{i}c^{\dagger}_{j}\Delta^V_{ij}
+c_{j}c_{i}\Delta^{V*}_{ij}
-V^{-1}\Delta^{V*}_{ij}\Delta^V_{ij}
\bigg]\notag\\
=&
\sum_{\bf k}\bigg[
c^{\dagger}_{A\bf k}c^{\dagger}_{B-\bf k}\Delta^U_{AB}({\bf k})
+c_{A-\bf k}c_{B\bf k}\Delta^{U*}_{AB}({\bf k})
+c^{\dagger}_{A\bf k}c^{\dagger}_{B-\bf k}\Delta^V_{AB}({\bf k})
+c_{A-\bf k}c_{B\bf k}\Delta^{V*}_{AB}({\bf k})
\bigg]+\text{const},
\end{align}
where $\Delta^{U(V)}_{ij}=-\Delta^{U(V)}_{ji}\equiv \Delta^{U (V)}$, $A$ and $B$ are two sublattice indices, and $c_{i}=\frac{1}{\sqrt{N}}\sum_{\bf k}c_{A/B,\bf k}e^{i{\bf k}\cdot {\bf r}_i}$ depending on whether ${\bf r}_i$ is in the $A$ or $B$ sublattice.

\begin{align}
\Delta^U({\bf k})
&=
\Delta^U
\sum_{i=1}^3
\begin{pmatrix}
0
&e^{-i{\bf k}\cdot{\bf a}_i}\\
-e^{i{\bf k}\cdot{\bf a}_i}
&0
\end{pmatrix}\notag\\
&=-i\Delta^U
\bigg\{[\cos({\bf k}\cdot{\bf a}_1)
+\cos({\bf k}\cdot{\bf a}_2)
+\cos({\bf k}\cdot{\bf a}_3)]\sigma_y
+[\sin({\bf k}\cdot{\bf a}_1)
+\sin({\bf k}\cdot{\bf a}_2)
+\sin({\bf k}\cdot{\bf a}_3)]\sigma_x\bigg\}
\notag\\
\Delta^V({\bf k})
&=
-2i\Delta^V\sum_{i=1}^3
\begin{pmatrix}
\sin({\bf k}\cdot{\bf b}_i)
&
0
\\
0
&
\sin({\bf k}\cdot{\bf b}_i)
\end{pmatrix}\notag\\
&=
-2i\Delta^V
[\sin({\bf k}\cdot{\bf b}_1)
+\sin({\bf k}\cdot{\bf b}_2)
+\sin({\bf k}\cdot{\bf b}_3)].
\end{align}
$\Delta^U({\bf k})$ and $\Delta^V({\bf k})$ are in the $B_{2u}$ and $B_{1u}$ irreducible representations of the $D_{6h}$ symmetry group, respectively.
The $B_{2u}$ pairing has two nodes at $k_y=0$ because it is odd under $M_y$, while $B_{1u}$ pairing open the full gap.
Accordingly, the $B_{1u}$ pairing is energetically favored.

The corresponding BdG Hamiltonian for the $B_{1u}$ pairing is
\begin{align}
\label{eq:hex.BdG}
H
&=
-\mu({\bf k})\tau_z
+f_1({\bf k})\tau_z\sigma_x
+f_2({\bf k})\tau_z\sigma_y
+\Delta_1({\bf k})\tau_y
+\Delta_2({\bf k})\tau_y\sigma_x
+\Delta_3({\bf k})\tau_y\sigma_y,
\end{align}
where $\Delta_1({\bf k})=\Delta^V({\bf k})$, and $\Delta_2({\bf k})=\Delta_3({\bf k})=0$.
The BdG Hamiltonian is symmetric under
\begin{align}
C=\tau_xK,\quad
T=K,\quad
C_3=1,\quad
M_z=1,\quad
M_x=\tau_z,\quad
M_y=\sigma_x,\quad
P=\tau_z\sigma_x,
\end{align}
Table.~\ref{tab:hex.parity} shows that $\nu_2^{\rm BdG}=1$ for $\mu=0.2$, $t=1$, $\Delta^V=0.05$, i.e., we have a second-order topological superconductor.
\begin{table}[h]
\begin{tabular}{c|cccc}
TRIM	&	$(0,0)$	&	$(\pi,\pi/\sqrt{3})$		&	$(-\pi,\pi/\sqrt{3})$		&	$(0,2\pi/\sqrt{3})$\\
\hline
parity	&	$(+,+)$	&	$(-,-)$	&	$(-,-)$	&	$(-,-)$
\end{tabular}
\caption{
Parity eigenvalues of the occupied states of the BdG Hamiltonian in Eq.~\eqref{eq:hex.BdG}.
$\mu=0.2$, $t=1$, $\Delta^V=0.05$.
}
\label{tab:hex.parity}
\end{table}

\subsection{Kagome lattice}

\subsubsection{Normal state}

The nearest-neighbor tight-binding model on the Kagome lattice features one exactly flat band and other two bands crossing at the $K$ points, forming two Dirac points.
Superconductivity on the Kagome Lattice has been studied in the context of the large correlation effect due to the flat band.
Here, we consider the superconductivity of spin-polarized electrons at the filling near the Dirac points.

The nearest-neighbor tight-binding Hamiltonian is given by
\begin{align}
\hat{h}
=\hat{h}_{\mu}+\hat{h}_t
=
-\mu\sum_{i}c^{\dagger}_{i}c_{i}
-t\sum_{\braket{ij}}c^{\dagger}_{i}c_{j}.
\end{align}
In momentum space, we have
\begin{align}
h({\bf k})
&=
\begin{pmatrix}
-\mu		&-2t\cos k_1		&-2t\cos k_2\\
-2t\cos k_1	&-\mu			&-2t\cos k_3\\
-2t\cos k_2	&-2t\cos k_3		&-\mu
\end{pmatrix},
\end{align}
where
\begin{align}
k_1
&
={\bf k}\cdot{\bf a_1}
=\frac{1}{2}k_x,\notag\\
k_2
&
={\bf k}\cdot{\bf a_2}
=-\frac{1}{4}k_x+\frac{\sqrt{3}}{4}k_y,\notag\\
k_3
&
={\bf k}\cdot{\bf a_3}
=-\frac{1}{4}k_x-\frac{\sqrt{3}}{4}k_y.
\end{align}
${\bf a}_{i=1,2,3}$ connects neartest neighbor sites.
Let us introduce Gell-Mann matrices $\lambda_{i=0,\hdots 8}$ for notational convenience.
\begin{align}
\lambda_0
&=
\begin{pmatrix}
1&0&0\\
0&1&0\\
0&0&1
\end{pmatrix},
\quad
\lambda_1
=
\begin{pmatrix}
0&1&0\\
1&0&0\\
0&0&0
\end{pmatrix},
\quad
\lambda_2
=
\begin{pmatrix}
0&-i&0\\
i&0&0\\
0&0&0
\end{pmatrix}\notag\\
\lambda_3
&=
\begin{pmatrix}
1&0&0\\
0&-1&0\\
0&0&0
\end{pmatrix},
\quad
\lambda_4
=
\begin{pmatrix}
0&0&1\\
0&0&0\\
1&0&0
\end{pmatrix},
\quad
\lambda_5
=
\begin{pmatrix}
0&0&i\\
0&0&0\\
-i&0&0
\end{pmatrix}\notag\\
\lambda_6
&=
\begin{pmatrix}
0&0&0\\
0&0&1\\
0&1&0
\end{pmatrix},
\quad
\lambda_7
=
\begin{pmatrix}
0&0&0\\
0&0&-i\\
0&i&0
\end{pmatrix}
\quad
\lambda_8
=\frac{1}{\sqrt{3}}
\begin{pmatrix}
1&0&0\\
0&1&0\\
0&0&-2
\end{pmatrix}.
\end{align}
Using Gell-Mann matrices, we can write the Hamiltonian as
\begin{align}
h({\bf k})
=-\mu\lambda_0-2t(\cos k_1\lambda_1+\cos k_2\lambda_4+\cos k_3\lambda_6).
\end{align}
Two Dirac points appear at $K=(\pi/\sqrt{3},\pi)$ and $K'=(-\pi/\sqrt{3},\pi)$ points at $E=-t/2-\mu$.
One can see that it has the form of Eq.~\eqref{eq:3N} with $\mu_1=\mu$, $\mu_2=\mu_3=0$, and $f_{i=1,2,3}=2t\cos k_{i=1,2,3}$.
The Hamiltonian is symmetric under
\begin{align}
T
=K,\quad
C_3
=
\begin{pmatrix}
0&0&1\\
1&0&0\\
0&1&0
\end{pmatrix},\quad
M_z
=1,\quad
M_x
=
\begin{pmatrix}
0&1&0\\
1&0&0\\
0&0&1
\end{pmatrix},\quad
M_y
=
\begin{pmatrix}
0&1&0\\
1&0&0\\
0&0&1
\end{pmatrix},\quad
P
=1,
\end{align}
where $C_3$ is the $120^{\circ}$ rotation about the $z$ axis, and $M_{i=x,y,z}$ is the mirror operation that flips the $i=x,y,z$ coordinate.

\subsubsection{Superconducting state}

Let us add Coulomb interactions.
Since there is no on-site Coulomb interaction due to the Fermi statistics of spin-polarized fermions, we consider the nearest-neighbor interaction only.
\begin{align}
H_{\rm int}
&=
-U\sum_{\braket{i,j}}n_in_j
\notag\\
&\approx 
\sum_{\braket{i,j}}\bigg[
c^{\dagger}_{i}c^{\dagger}_{j}\Delta_{ij}
+c_{j}c_{i}\Delta^{*}_{ij}
-U^{-1}\Delta^{*}_{ij}\Delta_{ij}
\bigg]
\notag\\
&=
\sum_{\bf k}\bigg[
c^{\dagger}_{A\bf k}c^{\dagger}_{B-\bf k}\Delta_{AB}({\bf k})
+c_{A-\bf k}c_{B\bf k}\Delta^{U*}_{AB}({\bf k})
\bigg]+\text{const},
\end{align}
where $A,B=1,2,3$ are sublattice indices, and
\begin{align}
\Delta({\bf k})
&=
-2i\Delta
\begin{pmatrix}
0			
&\sin({\bf k}\cdot{\bf a}_1)	
&\sin({\bf k}\cdot{\bf a}_2)
\\
\sin({\bf k}\cdot{\bf a}_1)	
&0				
&\sin({\bf k}\cdot{\bf a}_3)
\\
\sin({\bf k}\cdot{\bf a}_2)	
&\sin({\bf k}\cdot{\bf a}_3)
&0\
\end{pmatrix}\notag\\
&=
-2i\Delta
(\sin k_1\lambda_1+\sin k_2\lambda_4+\sin k_3\lambda_6).
\end{align}
This pairing matrix belongs to the $B_{1u}$ irreducible representation of the $D_{6h}$ group, i.e., it is invariant under $C_{3z}$ and has parity $(-,+,+)$ under mirror $(M_x,M_y,M_z)$.
It opens the full gap on the Fermi surfaces.
The BdG Hamiltonian has the form
\begin{align}
\label{eq:kgm.BdG}
H({\bf k})
&=
-\mu\tau_z\lambda_0-2t(\cos k_1\tau_z\lambda_1+\cos k_2\tau_z\lambda_4+\cos k_3\tau_z\lambda_6)\notag\\
&+2\Delta\sin k_1\tau_y\lambda_1
+2\Delta\sin k_2\tau_y\lambda_4
+2\Delta\sin k_3\tau_y\lambda_6.
\end{align}
It corresponds to Eq.~\eqref{eq:6BdG} with $\mu_1=\mu$, $f_{i=1,2,3}=-2t\cos k_{i=1,2,3}$, $\mu_2=\mu_3=\Delta_1=\Delta_2=\Delta_3=0$, and $\Delta_{i=4,5,6}=2\Delta\sin k_{i=4,5,6}$.

\begin{table}[b]
\begin{tabular}{c|cccc}
TRIM	&	$(0,0)$	&	$(\pi,\pi/\sqrt{3})$		&	$(-\pi,\pi/\sqrt{3})$		&	$(0,2\pi/\sqrt{3})$\\
\hline
parity	&	$(+,-,-)$	&	$(+,+,-)$	&	$(+,+,-)$	&	$(-,-,+)$
\end{tabular}
\caption{
Parity eigenvalues of the occupied states of the BdG Hamiltonian in Eq.~\eqref{eq:kgm.BdG}.
$t=1$, $\mu=-0.3$, and $\Delta=0.05$.
}
\label{tab:kgm.parity}
\end{table}

Table.~\ref{tab:kgm.parity} shows the parity eigenvalues of the occupied states of the BdG Hamiltonian for $t=1$, $\mu=-0.3$, and $\Delta=0.05$.
Although we introduce odd-parity pairing in a doped Dirac semimetal, we have $\nu_2^{\rm BdG}=0 \mod 2$.
This is because the unit cell in the Kagome lattice is not inversion-invariant (it is inversion-invariant only up to some lattice translation of sublattice sites), which is consistent with our formula Eq.~\eqref{eq:decomposition}.


%

\end{document}